\documentclass[lettersize,journal]{IEEEtran}

\usepackage{amsmath,amsfonts}
\usepackage{array}
\usepackage[caption=false,font=normalsize,labelfont=sf,textfont=sf]{subfig}
\usepackage{textcomp}
\usepackage{stfloats}
\usepackage{url}
\usepackage{verbatim}
\usepackage{graphicx}
\usepackage[square, comma, sort&compress, numbers]{natbib} 
\usepackage[linesnumbered,lined,ruled,vlined]{algorithm2e}
\usepackage{setspace}
\usepackage{xcolor}
\usepackage{color}
\usepackage{amsmath,epsfig,amssymb,bm,dsfont}
\usepackage{makecell}
\usepackage{pdfpages}
\usepackage{colortbl}
\usepackage{mathrsfs}
\usepackage{tikz}
\usepackage{pgfplots}
\usepackage{tabularx}
\usepackage{booktabs}
\usepackage{float}
\usepackage{multirow}
\usepackage{threeparttable}
\usepackage{pifont}
\usepackage{subcaption}  
\usepackage{soul}
\usepackage{hyperref}
\definecolor{mygray}{gray}{.92}
\usepackage{bbding}
\usepackage{wrapfig}
\usepackage{wasysym}

\SetKwProg{ParFor}{parallel for}{}{}

\hypersetup{
  colorlinks=true,
  linkcolor=black,
  urlcolor=black,
  citecolor=black
}

\begin{document}

\title{Designing Spatial Architectures for Sparse Attention: STAR Accelerator via Cross-Stage Tiling}

\author{Huizheng~Wang, 
        Taiquan Wei, 
        Hongbin Wang,
        Zichuan Wang,
        Xinru Tang, 
        Zhiheng Yue,~\IEEEmembership{Student Member,~IEEE,}
        Shaojun Wei,~\IEEEmembership{Fellow,~IEEE,}
        Yang Hu,~\IEEEmembership{Senior Member,~IEEE,}
        Shouyi Yin,~\IEEEmembership{Fellow,~IEEE} 

\thanks{
This work was supported in part by the National Science and Technology Major Project under Grant 2022ZD0115200; in part by the NSFC under Grant 62125403, Grant U24A20234, Grant 92464302 and Grant U24B20164; in part by the Beijing S\&T Project Z251100008425010; in part by Shanghai Municipal Science and Technology Major Project; the Natural Science Foundation of Jiangsu Province Basic Research Program under Grant BK20243042; in part by the Beijing National Research Center for Information Science and Technology; in part by the Northern IC Technology Innovation Center (Beijing) Co., Ltd under Grant QYJS20232801B; and in part by the Beijing Advanced Innovation Center for Integrated Circuits. An earlier version of this paper was presented at the  IEEE 57th Annual IEEE/ACM International Symposium on Microarchitecture (MICRO), 2024 [DOI: 10.1109/MICRO61859.2024.00093]. (Corresponding author: Yang Hu, email: hu\_yang@tsinghua.edu.cn).}

\thanks{Huizheng Wang, Taiquan Wei, Hongbin Wang, Zichuan Wang, Xinru Tang, Zhiheng Yue, Shaojun Wei, and Yang Hu are with the School of Integrated Circuits, Tsinghua University, Beijing, 100084, China.}

\thanks{Shouyi Yin is with the School of Integrated Circuits, Tsinghua University, Beijing, 100084, China, and Shanghai AI Lab, Shanghai, 200232, China.} 

}

\markboth{}%
{Huizheng Wang \MakeLowercase{\textit{et al.}}: Designing Spatial Architectures for Sparse Attention: STAR Accelerator via Cross-Stage Tiling}

\maketitle

\begin{abstract}
Large language models (LLMs) rely on self-attention for contextual understanding, demanding high-throughput inference and large-scale token parallelism (LTPP). Existing dynamic sparsity accelerators falter under LTPP scenarios due to stage-isolated optimizations. Revisiting the end-to-end sparsity acceleration flow, we identify an overlooked opportunity: cross-stage coordination can substantially reduce redundant computation and memory access. We propose STAR, a cross-stage compute- and memory-efficient algorithm–hardware co-design tailored for Transformer inference under LTPP. STAR introduces a leading-zero-based sparsity prediction using log-domain add-only operations to minimize prediction overhead. It further employs distributed sorting and a sorted updating FlashAttention mechanism, guided by a coordinated tiling strategy that enables fine-grained stage interaction for improved memory efficiency and latency. These optimizations are supported by a dedicated STAR accelerator architecture, achieving up to $9.2\times$ speedup and $71.2\times$ energy efficiency over A100, and surpassing SOTA accelerators by up to $16.1\times$ energy and $27.1\times$ area efficiency gains. Further, we deploy STAR onto a multi-core spatial architecture, optimizing dataflow and execution orchestration for ultra-long sequence processing. Architectural evaluation shows that, compared to the baseline design, Spatial-STAR achieves a $20.1\times$ throughput improvement.   
\end{abstract}

\begin{IEEEkeywords}
Transformer, attention sparsity, FlashAttention, top-k, tiling, distributed attention, spatial architecture.  
\end{IEEEkeywords}

\section{Introduction}\label{sec:intro}
\IEEEPARstart{E}{mpowered} by \emph{self-attention}, large language models (LLMs) have revolutionized fields such as chatbots \cite{radford2019language} and code generation \cite{achiam2023gpt}. The \textit{self-attention} processes three matrices: $\mathbf{Q}$ (query), $\mathbf{K}$ (key) and $\mathbf{V}$ (value). First, the attention matrix $\mathbf{A}$$\in\mathbb{R}^{S\times S}$ is computed by $\mathbf{Q}$$\times$$\mathbf{K}^T$, where $S$ denotes the sequence length. The resulting matrix $\mathbf{A}$ is then passed through a softmax function for normalization before being multiplied by $\mathbf{V}$ to generate the final output.

\begin{figure}[t]
\centering\includegraphics[width=\linewidth]{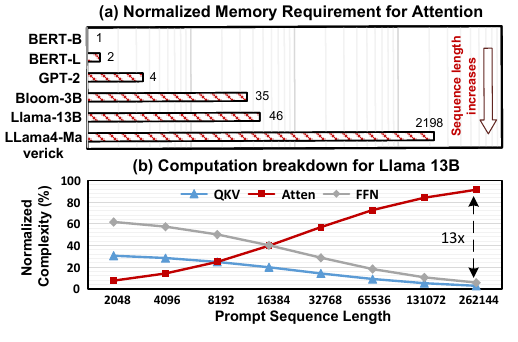}
\caption{(a) Normalized memory requirement for attention. (b) Computation breakdown for the Llama 13B.}
\label{fig:Fig1}
\end{figure}

{LLMs increasingly demand faster inference and higher throughput, particularly for long-context tasks. However, unlike the linear complexity of $O(SH^2)$ in the feed-forward network (FFN), where $H$ is the hidden dimension, the quadratic complexity $O(S^2H)$ of \emph{self-attention} severely hinders the efficiency of LLMs on long sequences. From early models such as BERT \cite{devlin2018bert} to recent ones like LLaMA\,4-Maverick \cite{meta2025llama4}, the maximum sequence length has expanded  over 32$\times$ (512 to 16k), whereas the hidden dimension $H$ has increased only 10$\times$ (768 to 8k). This dramatic growth of sequence length results in more than a $2000\times$ increase in attention memory footprint, as depicted in Fig.\ref{fig:Fig1} (a), creating significant barriers to deploying LLMs across both cloud and edge environments.  }

{Moreover, the quadratic computation of self-attention emerges as a critical bottleneck for fast inference. As shown in Fig.\ref{fig:Fig1} (b), when the sequence length reaches 16k tokens, attention surpasses the FFN as the most computation-intensive module. At 26k tokens, its cost escalates to nearly $13\times$ that of the combined QKV and FFN computations. These results underscore the pressing need to jointly optimize both computation and memory in self-attention.}

\begin{figure}[t]
\centering\includegraphics[width=0.95\linewidth]{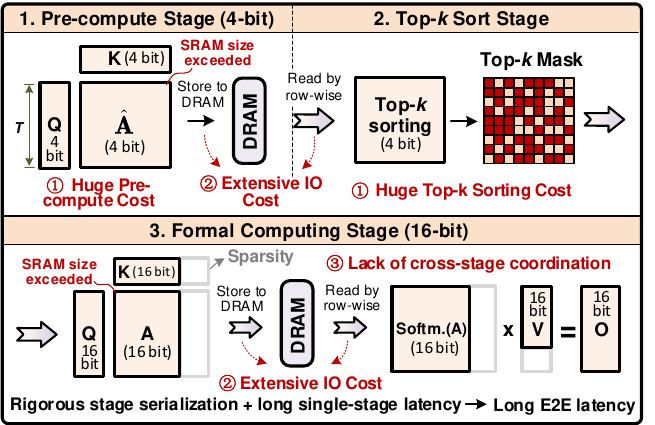}
\caption{{Challenges for scaling existing DS works to LTPP.}}
\label{fig:fig2}
\end{figure}

Owing to linguistic redundancy, not all contexts exhibit strong correlations, leaving attention with dynamic, data-dependent sparsity. To exploit it, \emph{Dynamic sparsity (DS) acceleration} \cite{ham20203,ham2021elsa,lu2021sanger,qu2022dota,qin2023fact,wang2021spatten,zhou2022energon} speculate vital Q-K pairs during runtime and calculate attention based only on these vital pairs, to evict unimportant computation. As depicted in Fig. \ref{fig:fig2}, DS consists of three stages: (1) \emph{Pre-compute stage}, estimating the attention matrix $\mathbf{\hat{A}}$ with low-precision arithmetic, e.g., 4-bit MSB. (2) \emph{Top-k stage}, retaining the most important $k$ Q-K pairs from each row. (3) \emph{Formal computing stage}, executing attention on the retained pairs with higher precision (e.g., 16-bit).

\textbf{\textit{Challenges:}} Limited on-chip SRAM confines most DS approaches to single-token serial processing, while modern LLMs demand support for ultra-long sequences. Large-scale token parallel processing (LTPP) reduces latency via parallelism and data reuse, necessitating DS accelerators capable of handling LTPP efficiently. However, scaling to LTPP presents three major challenges, as illustrated in Fig. \ref{fig:fig2}.

\begin{figure}[t]
\centering
\includegraphics[width=0.999\linewidth]{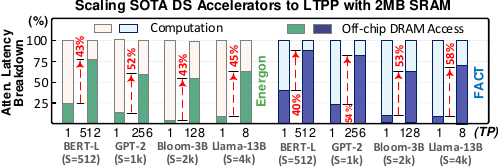}
\caption{Latency breakdown for SOTA DS accelerators (FACT \cite{qin2023fact}, Energon \cite{zhou2022energon}) with diverse token parallelism (TP).}
\label{fig:fig3}
\end{figure}

a) \emph{Prohibitive Computation Overhead}: The prediction stage (\emph{pre-compute stage} + \emph{top-k stage}), requires each query to multiply with the full-size K and perform sorting, without any sparsity reduction. For large token counts, the newly introduced stage incurs substantial computational and memory costs, which potentially offset the gains of sparsity acceleration. {A promising opportunity lies in transforming computation into the log-domain to eliminate costly multiplication.}

b) \emph{Unaffordable IO Cost $\&$ Latency}: The row-wise dependencies of \emph{top-k sorting} and \emph{softmax} require completing an entire row from the preceding matrix. This causes frequent off-chip DRAM accesses for intermediate data, becoming a major obstacle for current accelerators. As shown in Fig.~\ref{fig:fig3}, the memory access time (MAT) of two state-of-the-art (SOTA) DS accelerators rises sharply with token parallelism, averaging $72\%$ of total latency and becoming the main performance bottleneck. {A potential solution is to break down the row-wise dependency of top-$k$ sorting and softmax, by exploiting data locality and applying equivalent mathematical transformations.}

c) \emph{Uncoordinated workflow}: Existing DS accelerators lack cross-stage coordination, missing the chance to simplify later-stage operations through early-stage guidance. Despite FlashAttention (FA) \cite{dao2022flashattention} pioneering a tiling framework for \emph{softmax} to reduce memory access, it incurs extra computations from repeated exponentiation and comparison needed to update MAX values across tiles. Fortuitously, we identify an opportunity to simplify FA by utilizing top-$k$ information from the prediction stage. These challenges highlight the pressing need for more advanced DS strategies under LTPP scenarios.

To this end, this paper presents STAR, a cross-stage, synergic tiling algorithm-hardware co-design for attention optimization in LTPP scenarios. The contributions include:

    (1) The computation overhead in the \emph{pre-compute} stage is mitigated by a multiplier-free \emph{differential leading zero scheme (DLZS)}, which reduces sparsity prediction overhead. 
    
    (2) A \emph{sphere-search-aided distributed sorting (SADS)} strategy, which divides a long segment into sub-segments for individual sorting, effectively lowering comparison counts while enabling the tiling execution for top-$k$ stage.

    (3) A \emph{sorted updating FlashAttention (SU-FA)}, which decouples softmax row dependencies to enable tiling in the formal computation stage, while utilizing sorting information from the top-$k$ stage to reduce computation overhead.
    
    (4) A tailored accelerator is defined to support the above optimizations. Evaluated on $20$ benchmarks, STAR achieves an average energy efficiency of $7183$ GOPS/W, which is $71.2\times$ and average $8.6\times$ higher than Nvidia A100 GPU and three SOTA accelerators, respectively.
    
    (5) Based on STAR, we design a spatial architecture for distributed processing of ultra-long sequences. We propose a dedicated dataflow, \emph{DRAttention}, to minimize communication overhead, and a logic-level communication algorithm, \emph{MRCA}, to efficiently implement it under 2D mesh topology. The notations used throughout this paper are listed in Table \ref{tab:notation}.

\begin{table}[b]
\small
\renewcommand{\arraystretch}{1.05}
\centering
\caption{{List of notations used in this paper.}}
\begin{tabular}{ll}
\specialrule{0.12em}{0.5pt}{0.8pt}
\textbf{Notation} & \textbf{Description} \\
\specialrule{0.12em}{0.2pt}{0.8pt}
$S$; $H$      & Sequence length; Hidden dimension \\
$N_h$; $d_h$    & Number of heads; Hidden dimension of each head \\
$T$      & Number of queries to be processed in parallel \\
$T_r$; $T_c$    & Tile number along the row/column dimension \\
$B_r$; $B_c$    & Tile size along the row/column dimension \\
$\mathbf{Q}; \mathbf{K}; \mathbf{V}$ & Query, Key and Value matrices \\
$\mathbf{A}; \mathbf{\hat{A}}$ & Attention and estimated attention matrices \\
\specialrule{0.12em}{0.5pt}{1pt}
\end{tabular}
\label{tab:notation}
\end{table}

\section{Background and Motivation}\label{sec:background}
\subsection{Preliminaries for Transformer}\label{subsec:Transformer}
\begin{figure}[t]
\centering
\includegraphics[width=0.999\linewidth]{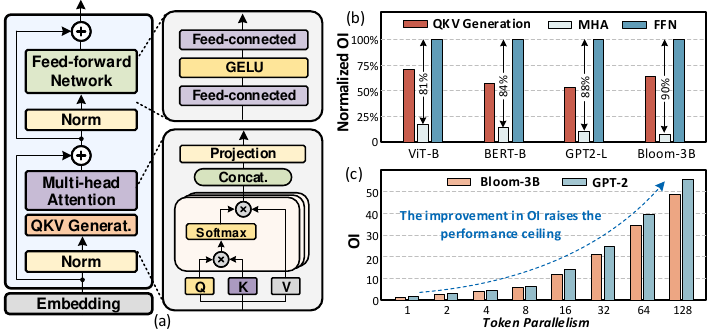}
\caption{(a) Transformer model. (b)  Operation intensity (OI) of diverse modules. (c) OI of MHA versus token parallelism.}
\label{fig:Transformer}
\end{figure}

Fig. \ref{fig:Transformer} illustrates the Transformer architecture and its key components. Initially, the input sequence of $S$ tokens is mapped into an embedding matrix $\mathbf{X}\in \mathbb{R}^{S\times H}$, where $H$ is the hidden dimension. Then, this input matrix $\mathbf{X}$ is projected into Q, K and V spaces, each of dimension $\mathbb{R}^{S\times H}$. Next, Q, K and V are split into {$N_h$ chunks}, with each chunk having a hidden dimension of {$H/N_h$, i.e., $d_h$}. These chunks are sent to the multi-head Attention (MHA) part, where the Q and K are multiplied to generate an attention matrix $\mathbf{A}\in\mathbb{R}^{S\times S}$, which represents the correlation of each Q-K pair. {The attention matrix is then normalized by the softmax operation as shown in Eq.~\eqref{eq:softmax1}. The softmax is applied in row-wise, and subtracting $\max_j (A_{i,j})$ from each row ensures numerical stability. }{
\begin{equation}
\text{Softmax}(A_{i,j})\!=\!\frac{e^{A_{i,j} - \max_j (A_{i,j})}}
{\sum_{k=1}^{S} e^{A_{i,k} - \max_j (A_{i,j})}}, j \!=\!1, 2, \dots, S
\label{eq:softmax1}
\end{equation}}

\noindent This normalized attention is multiplied with the V activations, resulting in a matrix {$\mathbf{O}\in\mathbb{R}^{S
\times d_h}$}, as depicted in Eq. \eqref{eq:attention}. Next, the outputs from all the attention heads are concatenated and projected by a weight matrix $\mathbf{W}_O\in \mathbb{R}^{H\times H}$. Finally, the FFN with two fully connected layers generates the final outputs.
\begin{equation}
\mathbf{O}={\rm softmax}\left(\mathbf{QK}^T/\sqrt{d_h}\right)\times \mathbf{V}, ~~{d_h=H/N_h}.
\label{eq:attention}
\end{equation}

We analyze the operational intensity (OI) \cite{williams2009roofline} of Transformer components, which reflects the intrinsic data reuse potential. As shown in Fig. \ref{fig:Transformer}(b), MHA exhibits the lowest computational intensity, only about $15\%$ of that of FFN. This suggests MHA demands significantly higher IO traffic per operation. Fig. \ref{fig:Transformer}(c) further shows that increasing token parallelism enhances the OI of MHA in both Bloom and GPT-2 models, as the Key matrix can be reused across more queries.

\subsection{Prohibitive Complexity for FlashAttention (FA)}\label{subsec:FlashAttention}
To mitigate the IO overhead of attention, the FA series \cite{dao2022flashattention,dao2023flashattention} adopt a tiling strategy that avoids materializing the full attention matrix on chip, thereby reducing costly SRAM–HBM data transfers. However, despite improved I/O efficiency, tile-wise incremental processing introduces substantial computational overhead. To quantify this, we profile the increased operations versus sequence length $S$, by fixing the tile size at $B_c$$=$$16$, i.e., $T_c$$=$$S/16$. As depicted in Fig.\ref{fig:Flash_Cost} (b), when $S$=$2048$, FA-2 consumes 8 million more exponentiations and 0.3 million more comparisons than the vanilla baseline. Since each non-matmul FLOP is roughly 16$\times$ more costly than a matmul FLOP \cite{dao2023flashattention}, this extra overhead severely impedes attention efficacy gains. 

Further, Fig.~\ref{fig:Flash_Cost}(c) presents the increase in computation complexity after aggregating all operations. {The total complexity, which includes different operation types, is unified by normalization to equivalent additions} \footnote{{We normalize the total computation complexity of attention with equivalent additions: $C_{\text{attention}} = 
\alpha \cdot N_{\text{add}} 
+ \beta \cdot N_{\text{mul}} 
+ \gamma \cdot N_{\text{cmp}} 
+ \delta \cdot N_{\text{div}} 
+ \epsilon \cdot N_{\text{exp}},$ where $\alpha = 1,\ \beta = 3,\ \gamma = 1,\ \delta = 8,\ \epsilon = 25$, according to the \cite{brent2010modern}.}}. As can be seen, the increased complexity escalates rapidly with increasing sequence length, with the extent of increase strongly correlating with the number of tiles $T_c$. A larger $T_c$ results in a more pronounced rise in complexity, driven by the higher frequency of redundant operations across tiles,  as illustrated in lines 5-8 of Fig.~\ref{fig:Flash_Cost}(a).

\subsection{Dynamic Sparsity in Transformer Attention}\label{subsec:Sparsity_in_attention}

While \textit{self-attention} is designed to capture token correlations within a context, not all tokens exhibit strong correlations. For example, tokens like articles `a' or `the' contribute negligibly to semantics and yield near-zero attention values. The subsequent \textit{softmax} would further suppress these small elements to near-zero. Therefore, their corresponding K/V vectors have a negligible impact on the attention output $\mathbf{O}$ and can be pruned with minimal performance loss.

To exploit this runtime sparsity of attention, a large body of DS accelerators \cite{ham20203,ham2021elsa,wang2021spatten,lu2021sanger,zhou2022energon,qin2023fact,qu2022dota} has been proposed. {Unlike static pruning techniques where the sparsity pattern is fixed offline, dynamic sparsity requires on-the-fly decisions during inference. The fundamental principle is to dynamically predict the most relevant Q–K pairs at runtime using an efficient, low-overhead mechanism and perform the high-precision attention computation only on the selected pairs. The general flow of DS is illustrated in Fig.~\ref{fig:fig2}. 


\section{Motivation}\label{sec:motivation}

\subsection{Challenges of LTPP with Current DS Accelerators}\label{subsec:Analysis_LTPP} 

In recent years, LLMs have seen an exponential increase in context length, from BERT’s 512 tokens in 2018 \cite{devlin2018bert} to GPT-4’s 32k tokens \cite{achiam2023gpt}, recently. This trend highlights the need for efficiently supporting LTPP. However, three key challenges lie in scaling current DS accelerators to LTPP scenarios: 

\begin{figure}[t]
\centering\includegraphics[width=0.999\linewidth]{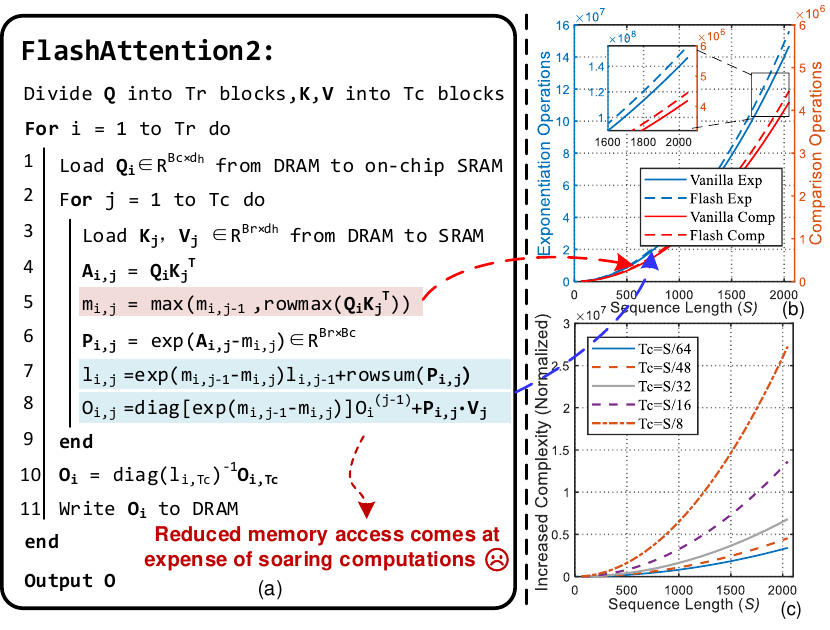}
\caption{{FlashAttention-2 (FA2) and its computation overhead.}}
\label{fig:Flash_Cost}
\end{figure}


{(1) \textbf{Prohibitive sparsity prediction overhead}. When $T$ queries are processed in parallel, the precomputation and sorting complexities scale as $\mathcal{O}(TSd_h)$ and $\mathcal{O}(TS^2k)$, where $k$$\in$$[0,1]$ denotes the top-$k$ ratio. Specifically, the pre-computation stage involves multiplying the Q matrix $\in$$[T,d_h]$ with the K matrix $\in$$[d_h,S]$, resulting in a complexity of $O(TSd_h)$. In the top-$k$ stage, the attention matrix $\in$$[T,S]$ is generated, where each row selects the top $S$$\cdot$$k$ elements. Since selecting each element requires $O(S)$ operations, the overall comparison and selection process incurs $O(TS^{2}k)$ complexity, resulting in significant runtime overhead.} For example, applying DS to LLaMA-7B with $T$$=$$512$ and $k$$=$$0.25$ incurs over $2.6$$\times$$10^8$ FLOPs and $2.1$$\times$$10^9$ comparisons for sparsity prediction. This results in nearly $12\times$ higher power overhead than the formal computation stage, when implemented on $45$\,nm CMOS at 500 MHz frequency. In addition, sparsity prediction accounts for around $57\%$ of the total attention latency, significantly limiting LLM inference efficiency. 


\begin{figure}[t]
\centering\includegraphics[width=0.999\linewidth]{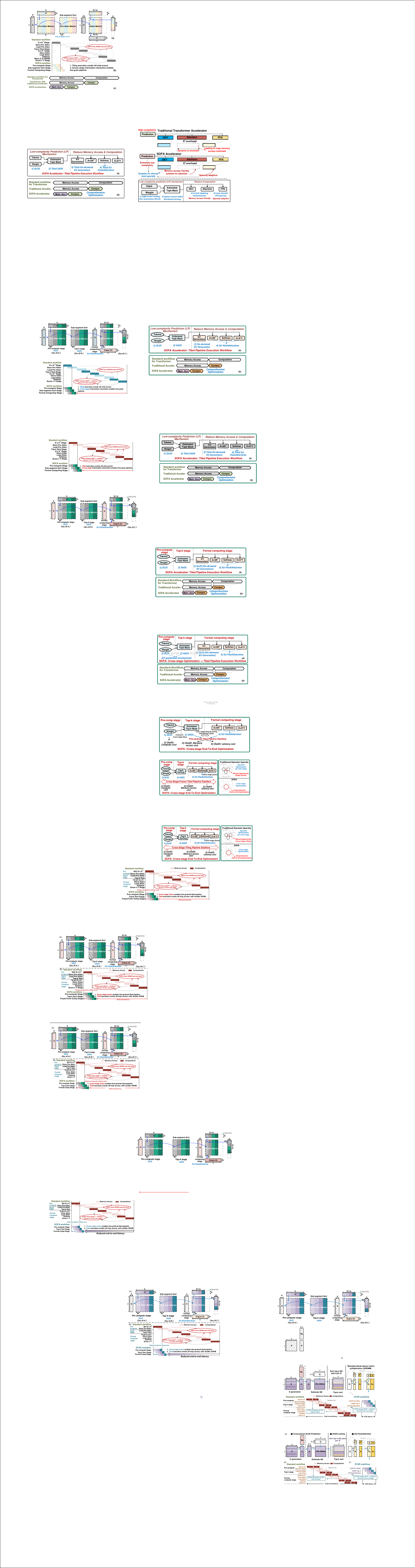}
\caption{End-to-end tiling pipeline workflow of STAR. (b) Comparison with the current standard DS workflow.}
\label{fig:SOFA_workflow}
\end{figure}

 (2) \textbf{Excessive IO overhead due to the rigidity of operator dependencies}. Due to the row-wise nature of the top-$k$ and softmax operators, directly increasing token parallelism leads to intermediate data exceeding the on-chip SRAM capacity. As a result, the data have to be offloaded to off-chip DRAM, incurring substantial IO transfer overheads. Taking CMOS $45$\,nm for an illustration, the energy cost for a DRAM access is around 5-20 pj/bit, which is orders of magnitude higher than on-chip SRAM (0.1pj/bit). In terms of bandwidth, off-chip memory (DDR4, 25.6 GB/s) typically exhibits two orders of magnitude lower bandwidth than on-chip memory (SRAM, 19TB/s). A direct but crude approach is to enlarge the on-chip SRAM capacity, but this would damage area efficiency. For example, when deploy Bloom7B with $T$$=$$512$, it requires a substantial 5MB of SRAM, resulting in $5.72$ mm$^2$ footprint under TSMC 28nm. This is $7.4\times$ and $8.9\times$ larger than the total area of SOTA ELSA \cite{ham2021elsa} and SpAtten \cite{wang2021spatten}, respectively.    

(3) \textbf{Overwhelming complexity due to lack of cross-stage optimization}. 
Based on online-softmax transformation, FA-2 employs a local tiling strategy to disaggregate the softmax into fragment operations. While the reduction in I/O complexity has been achieved, it comes with an intensive computational overhead, making it unsuitable for dynamic sparsity scenarios in LTPP. For example, when processing a sequence of length 1k with a tiling size of $B_c$$=$$4$, FA-2 must repeatedly perform exponentiation and comparison across blocks to ensure correctness of the global softmax result. This results in approximately $1.5\times$ more computation compared to the ideal implementation, resulting in a $3.37\times$ increase in computational energy consumption, due to the introduction of a significant number of expensive exponentiation operations.


In summary, the key bottleneck hindering DS accelerators in LTPP scenarios is the lack of cross-stage coordination, thus severely limiting memory and inference efficiency optimizations. Unfortunately, previous DS accelerators  \cite{ham20203,ham2021elsa,lu2021sanger,qu2022dota,wang2021spatten,zhou2022energon,qin2023fact} all overlook this critical aspect. A$^3$ \cite{ham20203}, ELSA \cite{ham2021elsa}, DOTA \cite{qu2022dota} and Sanger \cite{lu2021sanger} focus on alleviating the computing overhead of prediction stage, but all overlook the overhead introduced by memory accesses, let alone exploit the potential for cross-stage optimization. While SpAtten \cite{wang2021spatten} and Energon \cite{zhou2022energon} realize the challenge of off-chip memory access, SpAtten's head pruning causes irreversible accuracy degradation, and Energon's multi-round filter incurs prohibitive latency. Both therefore fail to handle the severe memory access overhead in LTPP scenarios. Overall, existing works focus on isolated, single-stage optimization and overlook cross-stage opportunities. This limitation restrains their ability to handle LTPP, which motivates our design of a cross-stage DS accelerator. 


\subsection{Extension for Spatial Architecture}
Although the monolithic STAR accelerator effectively exploits attention sparsity to achieve high computational efficiency, its single-core architecture faces fundamental scalability limitations. As LLMs rapidly expanding in both parameter size and sequence length \cite{kaplan2020scaling}, the demands on on-chip memory and computational resources increasingly surpass the capabilities of a single core \cite{hu2024wafer}. To overcome these challenges, multi-core spatial architectures are emerging as a promising solution for scalable LLM inference. Therefore, it is both significant and timely to investigate how the STAR accelerator can be effectively deployed within spatial architectures. 

To this end, we leverage the STAR as the basic computing core, exploring the multi-core partial architecture. This spatial architecture enables parallel execution across cores, distributed storage of key/value matrices, and reduced off-chip memory traffic through localized data reuse. However, the irregular and dynamic nature of sparse attention poses new challenges in inter-core communication and cross-core data orchestration. These challenges motivate our study of spatial dataflow and communication algorithms tailored to STAR-based multi-core architecture, aiming to enable scalable and efficient attention processing for ultra-large LLM workloads.

\begin{figure}[t]
\hspace{-1mm}
\includegraphics[width=\linewidth]{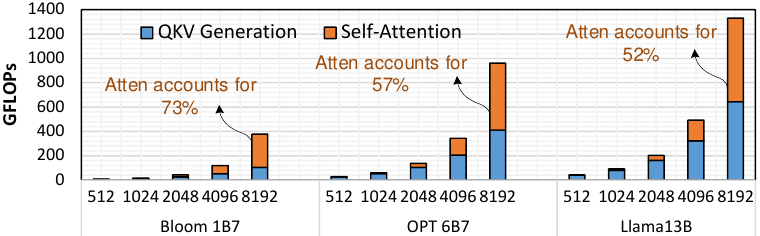}
\caption{Computational complexity between QKV and attention.}
\label{fig:QKV}
\end{figure}

\begin{figure*}[t]
\centering\includegraphics[width=\linewidth]{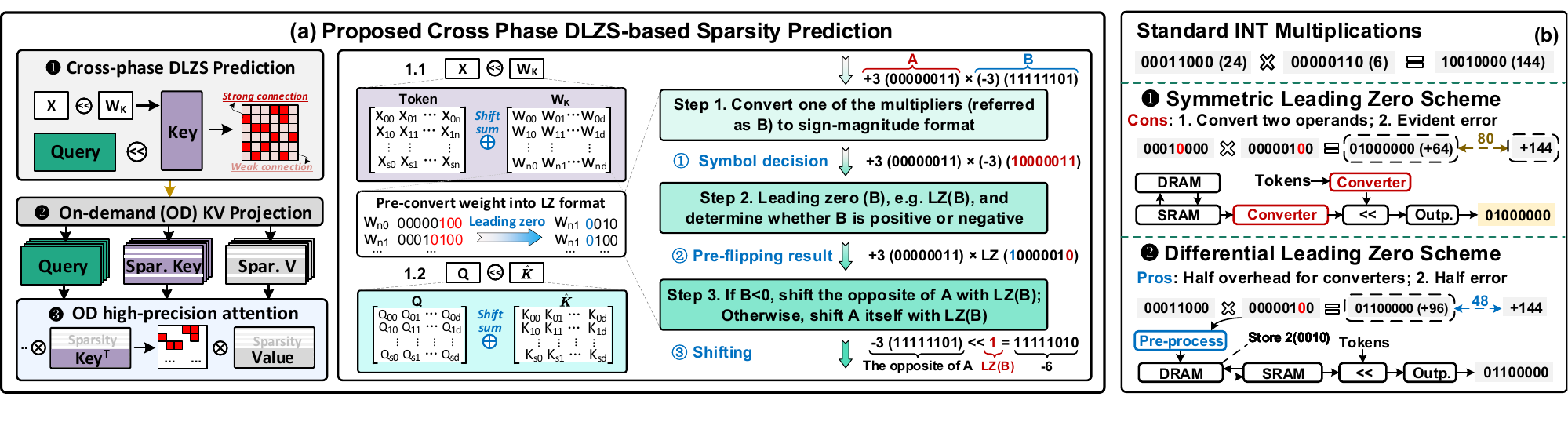}
\caption{(a) Cross-phase DLZS sparsity prediction and low-power PSP strategy in DLZS. {(b) Comparison of the proposed DLZS with the traditional symmetric leading zero scheme (SLZS) used in \cite{qin2023fact}.}}
\label{fig:DLOS}
\end{figure*}

\section{Optimizations for Cross-Stage Tiling}\label{sec:SOFA_algorithm}
Fig.~\ref{fig:SOFA_workflow} (a) summarizes three STAR optimizations addressing the challenges in Sect. \ref{subsec:Analysis_LTPP}. First, in the pre-computation stage, we introduce a cross-phase DLZS prediction scheme that is a multiplication-free method for estimating attention sparsity, while also extending the sparsity to KV generation. Second, we propose SADS, a distributed sorting strategy that partitions a long segment into multiple sub-segments, reducing sorting complexity and enabling cross-stage tiling. Third, we design SU-FlashAttention, which utilizes top-$k$ information to streamline the update operations in standard FlashAttention. Together, these techniques form STAR’s cross-stage tiling workflow, which alleviates off-chip memory access and end-to-end latency, as shown in Fig.~\ref{fig:SOFA_workflow} (b).

\subsection{Cross-Phase DLZS Sparsity Prediction}\label{subsection:DLZS}
Current DS works typically preserve dense QKV generation, leading to computation inefficiency. As shown in Fig. \ref{fig:QKV}, QKV generation can dominate the total cost for short sequences—exceeding attention when Bloom1B7 and OPT6B7 have fewer than 2k and 4k tokens, respectively. In contrast, attention becomes the bottleneck only beyond these lengths. Therefore, it is essential to jointly optimize QKV generation and self-attention to enhance overall efficiency.

To this end, we propose a cross-phase, on-demand sparsity mechanism. Instead of blindly generating all KVs, STAR only generates those KVs, whose attention scores are within the top-$k$ range, as illustrated in Fig. \ref{fig:DLOS}(a). However, it requires the \emph{pre-compute} stage first estimate $\mathbf{\hat{K}}$, based on which then speculates the $\mathbf{\hat{A}}$. Unfortunately, even utilizing low-precision multiplication (e.g. half-precision with MSBs only), it still incurs prohibitive power overhead. Therefore, a power- and accuracy-efficient prediction paradigm is essential.  
\begin{equation}
    x=Sign\times M\times2^{W-LZ},~~LZ\in\mathbb{Z}[1,\cdots, W].
    \label{eq:LoD1}
\end{equation}

Based on the log-domain transformation, we propose a multiplier-free paradigm, named \emph{differential leading zero scheme (DLZS)}. \emph{Differential} means: for a multiplication operation, only one operand undergoes a logarithmic-domain transformation to produce its leading-zero (LZ) format, in contrast to prior work \cite{qin2023fact} that applies logarithmic transformation to both operands. The underlying mathematical principle is as follows. For an INT-type number $x$, it can be represented by the Eq.~\eqref{eq:LoD1}, where $W$ is the quantized bitwidth, M stands for the mantissa lying within $[0,\,1]$, and LZ denotes its leading-zero value. Based on the representation, the multiplication of two INT numbers can be rewritten as Eq. \eqref{eq:LoD3}. By approximating $M_y$ as 1, we can derive the final computation formula as Eq. \eqref{eq:LoD4}. Since the bitwidth $W$ is fixed for a given model, DLZS directly shifts $x$ based on $LZ_y$ to estimate their multiplication, without expensive multipliers. 
\begin{subequations}
\begin{align}
&x\cdot y=\texttt{XOR}\left(S_x,S_y\right)M_x\cdot2^{(W-{LZ_x)}} M_y\cdot2^{(W-{LZ_y)}}\label{eq:LoD3}\\
&~~~~~~\approx\texttt{XOR}\left(S_x,S_y\right)M_x\cdot2^{(W-{LZ_x+W-LZ_y)}}\label{eq:LoD4}
\end{align}
\end{subequations}

However, it is non-trivial to achieve the \emph{differential} style log-domain computation. A key challenge is the bit flipping induced by sign determination after shifting, which results in prohibitive power overhead and may offset the benefits of log-domain approximation. To this end, we design a \emph{preflipping via symbol prediction (PSP)} strategy, as shown in Fig. \ref{fig:DLOS} (a) right. Initially, one operand $B$ is converted to a sign-magnitude representation. The LZ count of $B$, i.e., LZ(B), is then computed, followed by checking the sign of $B$. If $B$ is negative, the opposite of the other operand (that is $A$) is shifted. Otherwise, the operand $A$ itself is shifted. In this way, we can avoid the substantial power overhead from bit flipping. 

As depicted in Fig. \ref{fig:DLOS} (a), the  cross-phase DLZS prediction leverages a potential opportunity to further reduce the predicted power consumption. Specifically, since the weights are pre-known and keep unchanged during inference, DLZS pre-converts $\mathbf{W}_k$ into LZ format. This allows the first phase, that is \emph{Key prediction phase} (1.1), to directly shift the input based on the pre-converted LZ($\mathbf{W_k}$), without any extra LZ coding. In the subsequent \emph{Attention prediction phase} (1.2), to alleviate the inaccuracy accumulation, we apply LZ encoding for $\mathbf{Q}$, instead of $\mathbf{\hat{K}}$. Overall, compared to previous DS prediction strategies, the proposed cross-phase DLZS features lower conversion overhead and a wider range of sparsity acceleration. 

{As depicted in Fig. \ref{fig:DLOS} (b), compared to the traditional symmetric leading zero scheme (SLZS) used in FACT \cite{qin2023fact}, the proposed DLZS offers three key advantages: \textbf{a) Lower conversion overhead}. This is because DLZS performs the leading-zero conversion on only one operand, while the SLZS needs to process both. \textbf{b) Higher accuracy}. This is because the error of the leading-zero scheme originates from the loss of information carried by the bits following the most significant ‘1’. Since DLZS performs the leading-zero conversion on only one multiplier, it introduces less error and achieves higher precision. \textbf{c) Reduced memory access}. The reduction in memory access stems from the fact that DLZS only needs to load a 4-bit LZ value that has been pre-converted offline, whereas SLZS must load the full 8-bit operand.}

\subsection{Sphere-search Aided Distributed Sorting (SADS)}\label{subsec:SADS}
To identify important tokens from the estimated attention $\mathbf{\hat{A}}$, existing DS methods typically apply sorting or thresholding at the entire row level. However, these coarse-grained approaches incur significant computational overhead for long sequences, rendering them impractical for LTPP scenarios. 


\begin{figure}[t]
\centering\includegraphics[width=\linewidth]{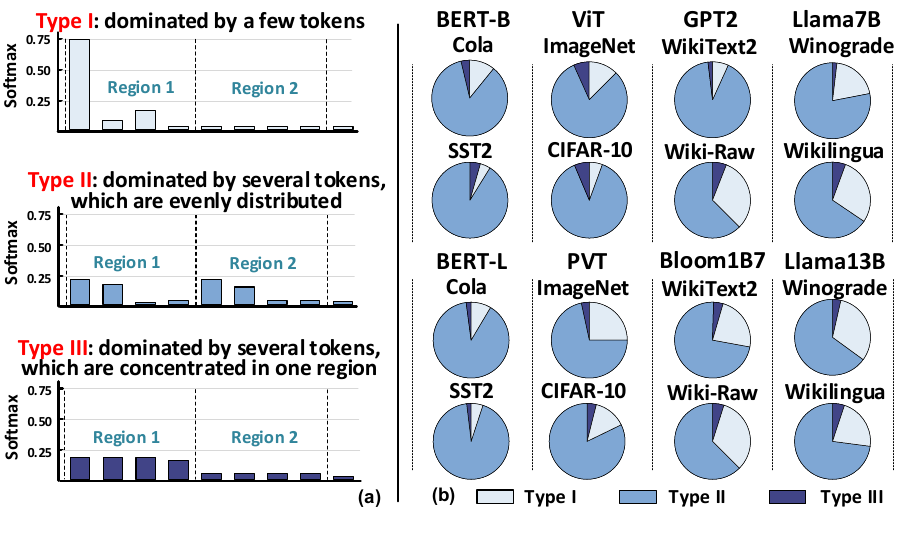}
\caption{{The data distribution of attention across diverse LLMs.}}
\label{fig:SADS_algorithm1}
\end{figure}

We begin by analyzing the softmax function and show that it it naturally supports lightweight max-based decision-making. Without loss of generality, consider a two-element vector $[x_0,x_1]$ processed by softmax, where $x_1$ is the max element and $x_1$$=$$x_0$$+$$\Delta$. As shown in Eq.\eqref{eq:softmax}, the softmax output for $x_0$ decays exponentially with its distance $\Delta$ from $x_1$. This observation implies that, after softmax, an element’s contribution is inversely correlated with its distance from the max: the farther it is, the smaller its impact on the final output.
\begin{equation}
\!\!\!x_1\!=\! x_0 \!+\! \Delta \Rightarrow \mathrm{softmax}(x_0) \!=\! \frac{e^{x_0}}{e^{x_0} \!+\! e^{x_0 + \Delta}}\!=\!\frac{1}{1 \!+\! e^{\Delta}}\!<\!\frac{1}{e^{\Delta}}\!
\label{eq:softmax}
\end{equation}

Since softmax exponentially amplifies the differences between elements, its output is mainly determined by the dominant tokens in the inputs. We classify the data distribution of each self-attention row into three types, as shown in Fig.\ref{fig:SADS_algorithm1} (a). Type I: A few highly dominant tokens; Type II: Larger tokens evenly distributed across different regions; Type III: Larger tokens concentrated in a specific region. {To characterize their distribution during inference, we analyze attention from several representative models and tasks, including BERT-B/L, ViT, PVT, GPT2, Bloom-1B7, Llama-7B, and Llama-13B. The analysis covers tasks spanning logic reasoning, text generation, and vision benchmarks (e.g., SST2, ImageNet, WikiText2, Winogrande), each based on 4k randomly sampled rows.} Across all scenarios, Type II dominates with an average proportion of about $73\%$. Meanwhile, Type I is more likely to occur in ViT, GPT, and LLaMA, with an average proportion of $22\%$, whereas in the encoder-based model BERT, it is only $12\%$. This discrepancy may be attributed to the local similarity features in images and the recent token characteristics inherent to auto-regressive decoder. By contrast, Type III occurs with negligible probability, which even approaches zero in both GPT-2 and LLaMA. This is likely due to the increased token length, where larger tokens tend to be distributed throughout the context, rather than being concentrated in a specific region.

\begin{figure}[t]
\centering\includegraphics[width=\linewidth]{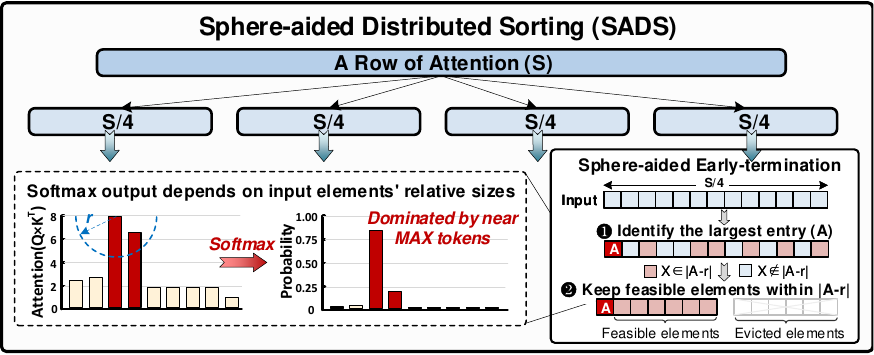}
\caption{{Sphere-search aided distributed sorting (SADS).}}
\label{fig:SADS_algorithm2}
\end{figure}

For Type I and Type II, which together account for more than $95\%$ of the total distribution, their primary commonality lies in their dispersed and relatively uniform attention distribution. This characteristic enables local maxima across different regions to be reasonably regarded as global maxima. Therefore, a strategy that can efficiently capture local maxima within Type I and Type II distributions in a cheap manner naturally represents a more effective token selection strategy.

To address this, we propose a \emph{distributed sorting strategy named SADS}, which also incorporates a concept analogous to a spherical radius to enable early termination of redundant sorting for trivial data. As illustrated in Fig. \ref{fig:SADS_algorithm2}, SADS first partitions an ultra-long sequence into $n$ sub-segments (e.g., $4$) for distributed sorting. Each segment then selects its own top-$k/n$ largest values. {Instead of naive sorting, SADS introduces a \textit{spherical radius-based pruning criterion} for early termination. Specifically, it first identifies the largest entry $A$ \ding{182} then defines a feasible region centered at $A$, denoted as $x\in\vert A-r\vert$. The subsequent top-$k/n$ sorting is restricted to elements in the feasible region \ding{183}. For an element $x$, if its distance to the max satisfies $\Delta>r$, it is eliminated. This is justified by the analysis in  Eq.~\eqref{eq:softmax}, which shows the softmax value of an element $x$ decreases exponentially with its distance $\Delta$ from the maximum. The radius $r$ is user-configurable and is empirically set to 5 in our experiments. Under this setting, the softmax value of elements outside the feasible region drops below 0.0067, indicating their negligible contribution to the attention output.}

{\textbf{Complexity Analysis}. Compared with standard sorting of complexity $O(SSk)$, SADS reduces it to $O(SSk\rho / n)$, where $k$ denotes the top-$k$ ratio ($0 < k \le 1$), $n$ is the number of sub-segments, and $\rho$ represents the remaining element ratio after early termination. In a typical setting with $S = 1024$, $n = 4$, $k = 0.25$, and $\rho = 0.4$ (with $r=5$), the complexity of SADS is only about 10\% of that of standard sorting. Notably, the specific number of sub-segments (e.g. tiling size) of each layer is required to collaborate with SU-FA and is obtained by the DSE in Appendix A.}


\subsection{Sorted-Updating FlashAttention (SU-FA)}\label{subsec:SU-FlashAttention}
Existing DS methods alleviate the computational burden of attention but fail to optimize memory access, whereas FA reduces memory access at the cost of non-matmul overhead. To this end, we propose SU-FA, which utilizes the top-$k$ information from DS to mitigate the max-value refreshes in FA, enhancing both computational and memory efficiency. 

As shown in lines 5 and 7 of Fig. \ref{fig:Flash_Cost}, redundant computations in FA mainly result from max-value updates across tiles and associated exponential operations. An intuitive approach to mitigate these inefficiencies is to leverage top-$k$ information, which helps reduce these redundant computations. One straightforward scheme is to directly compute the max score based on the estimated index and then perform the attention calculation using the conventional softmax. However, there are two critical issues: 1) Due to the approximation property of DLZS, the estimated max index may be inaccurate, potentially leading to numerical overflow. 2) The separate computation of the max score incurs additional computational burden. 

\begin{figure}[t]
\centering
\includegraphics[width=\linewidth]{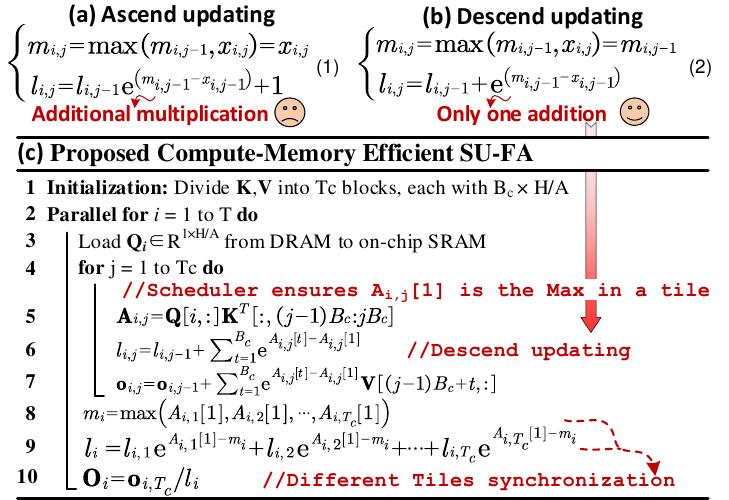}
\caption{{(a) Redundant operations in FA. (b) Formulas for diverse updating orders. (c) Procedure of SU-FA.}}
\label{fig:SU-FA}
\end{figure}

To address this, we propose a \emph{successive sorted-updating} strategy. \emph{Successive} means: it retains the same serial computation paradigm as FA, thus avoiding the overhead of separately computing the max score. \emph{Sorted} indicates that the updating process is performed in accordance with the order of top-$k$ indices. Naturally, there are two potential update orders: \textit{ascending} and \textit{descending}. This leads to the next question: which update order is more effective? Fig.\ref{fig:SU-FA} (b) presents the update formulas derived from top-$k$ simplification for different strategies. Though both strategies eliminate the comparisons for the max value, they still leads to discrepancies in multiplication overhead. Compared to descend updating, ascend updating results in an additional multiplication operation at each step. While this discrepancy may seem trivial, in long-sequence scenarios, it would be accumulated and transformed into prohibitive overhead. For instance, with an $8$\,k token sequence, ascend updating incurs $2.1$$\times$$10^6$ more multiplication operations, compared to descend updating. Therefore, SU-FA adopts the descend updating strategy by default.

To sum up, in the algorithm optimization of STAR, the granularity of sub-segment $S_i$ and top-$k$ jointly define an interesting design space. Smaller $S_i$ leads to higher SU-FA overhead due to synchronization across sub-segments and exacerbates performance degradation due to more fragmented computation, but reduces sorting complexity. By contrast, larger $S_i$ raises sorting overhead but lowers SU-FA overhead, and generally yields better accuracy. To this end, we present an efficient design space exploration (DSE) strategy to explore the optimal sub-segment size. Please refer to Appendix A.

\section{Architecture and Hardware Innovation}\label{sec:SOFA_Hardware}

\S\ref{subsec:Architecture_overview} introduces the overall STAR accelerator architecture, which effectively translates the theoretical algorithmic benefits into practical speedup. A detailed description of the primary hardware components is provided in Appendix B. Next, \S\ref{subsec:Spatial Extension} discusses the extension of STAR into the spatial architectures.

\subsection{Overall Architecture}\label{subsec:Architecture_overview}
Fig. \ref{fig:SOFA_Hardware} depicts the high-level architecture of STAR accelerator, which consists of six main components: a cross-stage DLZS prediction unit, an SADS sorting unit, a PE array for KV on-demand generation, an SU-FA execution unit, a tiled $\&$ and out-of-order scheduler, and on-chip SRAM. STAR is designed to process $128$ queries in parallel. First, the controller instructs the fetcher to load corresponding $\mathbf{Q}$, $\mathbf{X}$ and $\mathbf{W}_k$ into the on-chip SRAM \ding{202}. Then, the DLZS unit begins cross-stage prediction, which generates an estimated attention scores $\mathbf{\hat{A}}$ \ding{203}. Next, based on the $\mathbf{\hat{A}}$, the SADS unit identifies the indices of the corresponding key matrix for the top $k/n$ scores in each row \ding{204}. The indices are then sent to the scheduler \ding{205}, which generates an intermediate binary mask \ding{206} to guide the PE array in producing the necessary KVs \ding{207}. Once the KVs are generated on-demand \ding{208}, the SU-FA unit is triggered to execute the designed \textit{descend updating} workflow, as presented in Sec. \ref{subsec:SU-FlashAttention}. The final results are stored back to off-chip DRAM \ding{209}.   

\begin{figure}[t]
\centering
\includegraphics[width=0.99\linewidth]{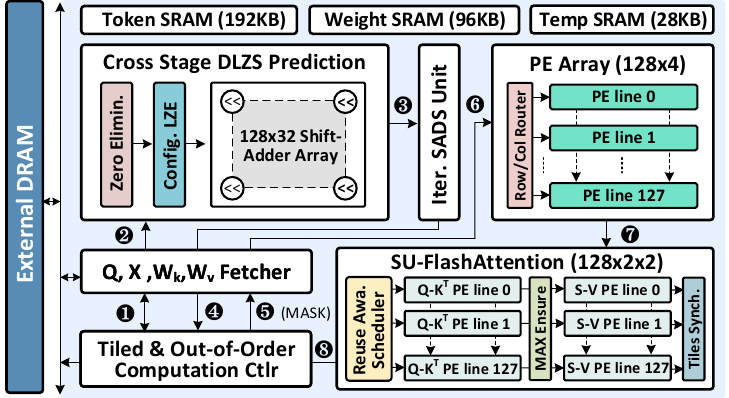}
\caption{High-level block diagram for the STAR accelerator.}
\label{fig:SOFA_Hardware}\vspace{-2mm}
\end{figure}

\subsection{Spatial Extension Design for STAR}\label{subsec:Spatial Extension}

Multi-chiplet integration offers a promising path to large-scale systems \cite{wang2025watos,wang2025temp}. Typically, it employs a 2D mesh topology-based spatial architecture, as mesh exhibits lower wiring costs and superior scalability compared to alternative topologies \cite{kandalla2010designing} such as Fat-tree, Dragonfly and Switch-leaf.

\begin{wrapfigure}{r}{0.255\textwidth}\vspace{-3mm}  
    \centering
    \includegraphics[width=0.25\textwidth]{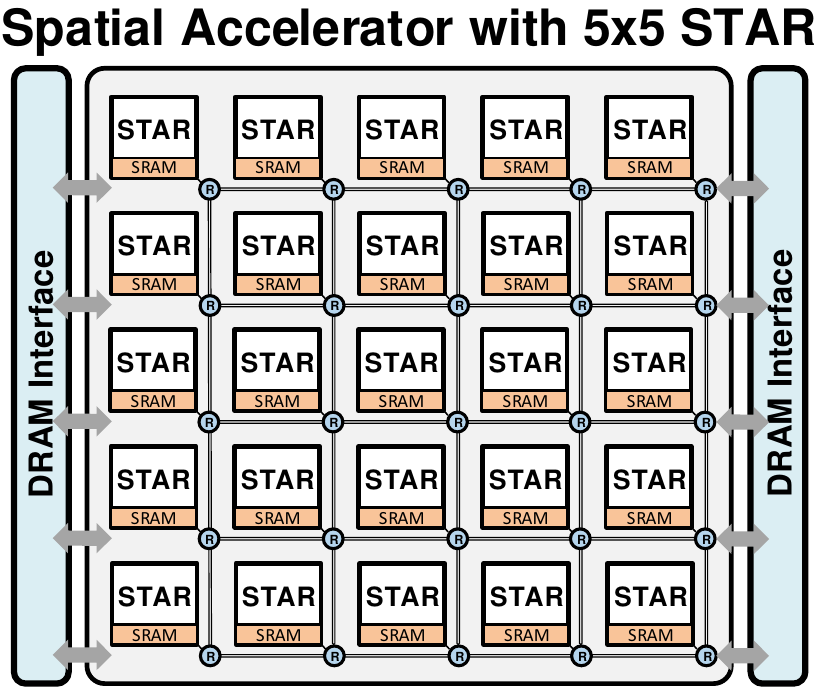} 
    \caption{5$\times$5 spatial archi. setup.}
    \label{fig:spatial_archi}\vspace{-2mm}
\end{wrapfigure}




As depicted in Fig. \ref{fig:spatial_archi}, we construct a 5×5 two-dimensional (2D) mesh spatial architecture, where each node instantiates the STAR accelerator as the fundamental computing unit. {Notably, the choice of a 5$\times$5 configuration represents a balanced trade-off between parallel scalability and communication overhead, while also following the design scale adopted in Dojo \cite{talpes2022dojo}.} Each unit comprises a private SRAM block for local data storage and a dedicated router supporting five-directional routing (north, south, east, west, and local), thereby enabling flexible and low-latency communication across the mesh. The computing units are interconnected via a 2D mesh Network-on-Chip (NoC), enabling efficient horizontal and vertical data transfers for parallel execution and dynamic workload balancing. DRAM modules are symmetrically placed on both sides of the mesh. Memory requests issued by compute units are transmitted over the NoC using dynamic routing, and the requested data is returned to the corresponding units via multi-hop paths.


However, achieving efficient attention parallelism on the 2D mesh spatial architecture is non-trivial and presents two primary challenges: First, at the dataflow level, it is essential to carefully orchestrate the communication paths of Q, K, and V to minimize global traffic while maintaining regular and predictable communication patterns. Second, at the communication-logic level, it must be meticulously designed to ensure that the logical data movement aligns precisely with the 2D mesh physical topology, for mitigating tail latency.

To address these challenges, in this section, we first introduce a dedicated distributed attention dataflow strategy termed \emph{DRAttention}. Then, to enable its effective realization on the 2D mesh physical topology, we further propose an optimized logical communication algorithm, referred to as \textit{MRCA}.

\subsubsection{Distributed Ring-flow-based Attention (DRAttention)}\label{subsec:Distributed_Attention}
Fig. \ref{fig:DRAttention} illustrates the data partitioning and computation scheduling strategy of DRAttention on the $5\times 5$ 2D mesh architecture. {Before computation, both the Query tensor $\in \mathbb{R}^{S\times d_h}$ and the Input tensor X $\in \mathbb{R}^{S\times H}$ are partitioned only along the sequence dimension, while the hidden dimension remains intact. The Query tensor is divided into 25 sub-blocks of size $\in \mathbb{R}^{S/25 \times d_h}$, which are distributed across both the rows and columns of the STAR array. Specifically, consecutive groups of five sub-blocks are mapped to the five STAR units within the same row, with each unit processing one sub-block. By contrast, the input tensor X is divided into only five sub-blocks of size $\mathbb{R}^{S/5\times H}$, which are mapped along the column dimension, with each sub-block shared by the five STAR units within the same column.}

Each STAR unit on-demand generates the required KV tensors corresponding to its current sub-block $X_i$. Once the computation begins, the STAR unit executes the sparse attention as depicted in Section \ref{subsec:Architecture_overview}. During this process, for any STAR unit, while computing attention between
its local Q sub-block and the generated KV tensors, the STAR unit concurrently transmits its Q sub-block to the next STAR unit and receives a Q sub-blocks from the preceding unit, as depicted in the Fig. \ref{fig:DRAttention} (b). If the computation time exceeds the time required for transferring the Q sub-blocks, no additional communication overhead is incurred. Further, certain intermediate data, such as the local maximum $m_i$ and partial sum $l_i$, are propagated alongside the Q sub-blocks. These values are incrementally updated at each time step to support normalization and maintain numerical stability. In this way, after the fifth time step, each STAR unit updates the attention result of its own Q sub-block using the global maximum value and generates the corresponding final attention results. 

\begin{figure}[b]
\centering
\includegraphics[width=\linewidth]{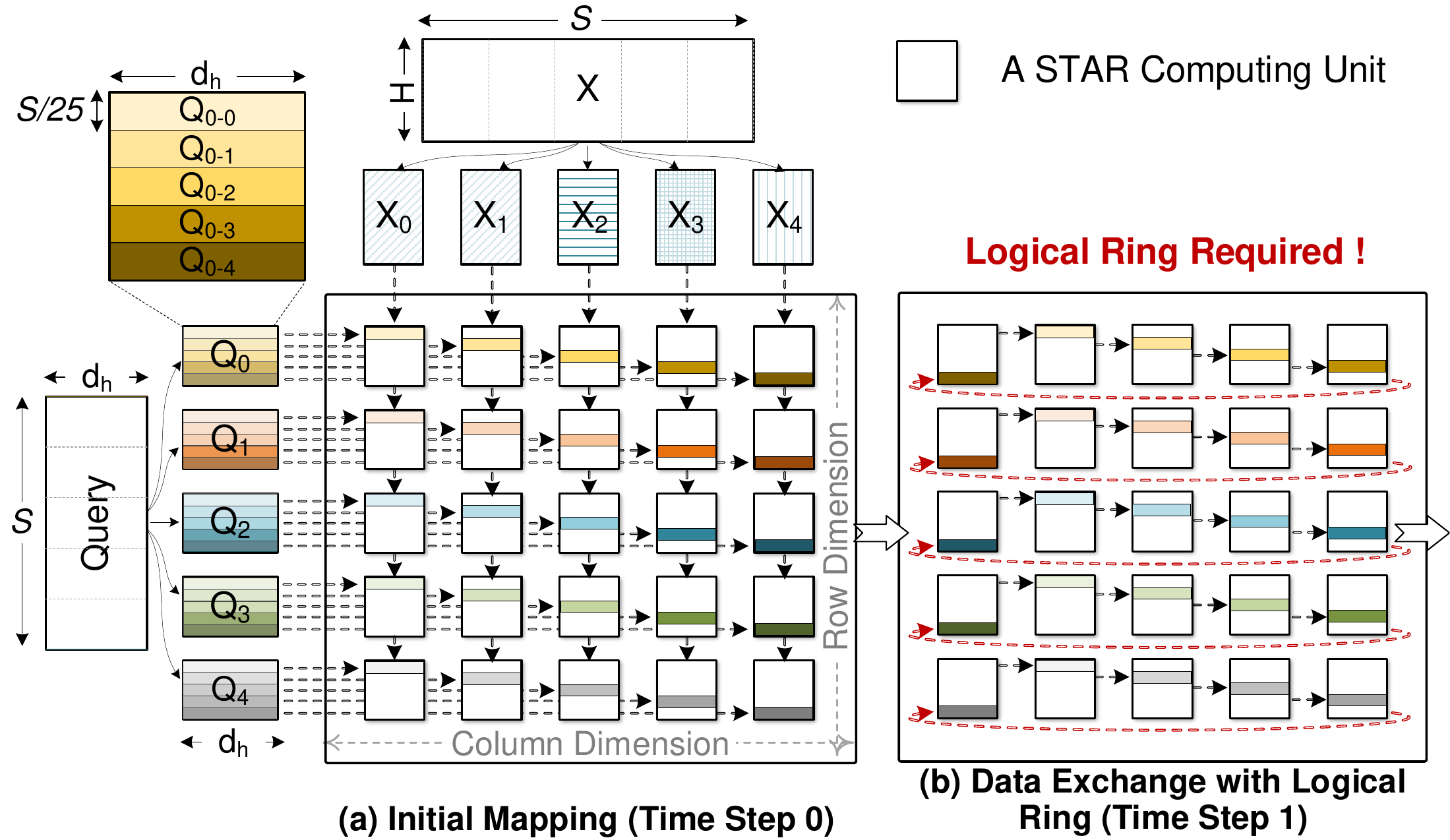}
\caption{{Tailored dataflow for DTAttention mechanism.}}
\label{fig:DRAttention}\vspace{-2mm}
\end{figure}

Overall, DRAttention offers two significant advantages. First, its Query-driven communication leads to low global communication overhead, as the Query tensor is substantially smaller than the K/V tensors. This reduction in communication volume helps alleviate NoC congestion. Second, DRAttention enables substantial overlap between computation and communication, effectively mitigating the potential performance degradation caused by exposed communication latency and improving compute unit utilization.

However, from the perspective of communication logic, DRAttention relies on a near-circular (a.k.a ring-style) data transfer pattern during execution, as illustrated by the red arrow in Fig. \ref{fig:DRAttention}(b). This logical communication behavior requires physical wrap-around links to enable cross-boundary data propagation. However, in practical hardware implementations, such a wrap-around link is generally infeasible due to physical constraints such as excessive routing distance and signal integrity issues. Therefore, a 2D mesh topology cannot, in practice, provide the degree of interconnectivity required to directly realize this communication pattern.

To resolve the mismatch between logical communication requirements and physical topology constraints, we propose an optimized communication scheduling algorithm, named \emph{MRCA}. This approach enables DRAttention to achieve the desired circular communication behavior within a standard mesh topology, without incurring additional physical interconnect costs, thereby facilitating efficient attention parallelism.


\subsubsection{Mesh friendly Ring Communication Algorithm (MRCA)}\label{subsec:Distributed_Attention}
Instead of forcing a logical ring topology onto a physical mesh topology, MRCA leverages the natural characteristics of the mesh to construct a congestion-free orchestration that is logically equivalent to a ring. The central innovation is inspired by the behavior of reflux tides, analogous to the reverse wave of a progressing wave. This mechanism, referred to as \textit{Reflux}, enables elegant transfer of the Q-stream and scales effectively to any number of computing units. The details of MRCA is summarized in Alg. \ref{alg:Q-stream}, while an execution example on a $1$$\times$$5$ spatial architecture (i.e., a 1D mesh with five computing units, CUs) is depicted in Fig. \ref{fig:Q stream}. For illustrative purposes, the 1D mesh is drawn in a vertical layout. 

\begin{figure}[t]
\centering
\includegraphics[width=0.95\linewidth]{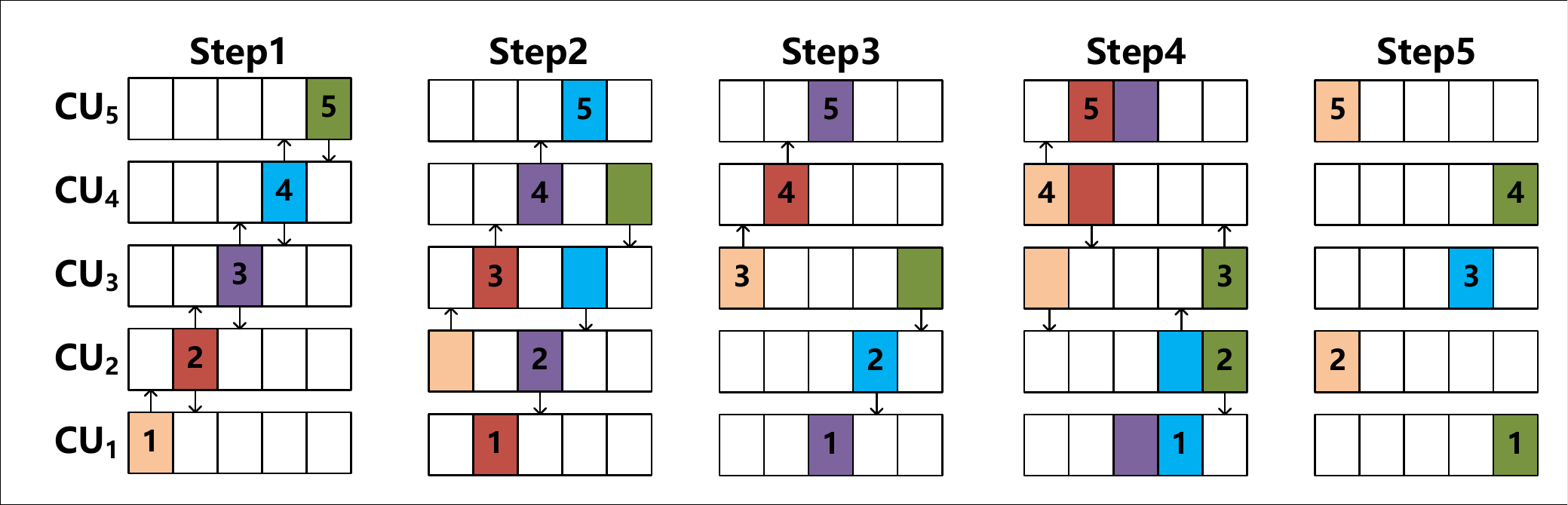}
\caption{Q stream in a $1$$\times$$5$ spatial accelerator. CU$_i$ denotes the computing unit with ID $i$. Different colors represent different data chunks. Blank chunks indicate that the corresponding chunk does not reside on the CU. The number $i$ indicates the chunks involved in the CU$_i$'s computation at the step.}
\label{fig:Q stream}\vspace{-4mm}
\end{figure}

In Alg. \ref{alg:Q-stream}, the communication mechanism is categorized into two types of operations: progress wave (lines 4-9) and reflux tides (lines 10-19). Progress wave occurs in nearly all $N$ (5) steps. The transferred Query sub-blocks (For short, we denote them as chunks) behave like waves spreading from the center outwards. We consider the transfer from a CU with a smaller ID to one with a larger ID as an upward wave (lines 4-6 in Alg. \ref{alg:Q-stream}), and the transfer from a CU with a larger ID to one with a smaller ID as a downward wave (lines 7-9).

As illustrated in Step2 of Fig. \ref{fig:Q stream}, following lines 4 and 7 in Alg. \ref{alg:Q-stream}, CU$_2$, CU$_3$, and CU$_4$ participate in both upward and downward wave transfers. Taking CU$_2$ as an example, in Step2, it stores chunk1 and chunk3. It forwards chunk1 to CU$_3$ (lines 5-6 in Alg. \ref{alg:Q-stream}) and simultaneously passes chunk3 downwards to CU$_1$ (lines 8-9). Therefore, the progress wave achieves the transmission of each chunk to the dies on both sides through the spread of chunks from the center, ensuring that each CU does not store more than 2 chunks per step.

The reflux tides occur after the $\lceil\frac{N}{2}\rceil$-th step, with the primary aim of ensuring that each CU can complete computations on different chunks within $N$ steps. To maintain computational balance, each CU would store at most two chunks in certain steps but would only compute one of them. For instance, consider CU$_2$ in Fig. \ref{fig:Q stream}, which stores chunk1 and chunk3 at Step\,2; CU$_2$ only utilizes chunk3 for computation in this step, neglecting chunk1. Without reflux tides, where chunk1 resides solely on CU$_2$ at Step\,2 due to progress waves, CU$_2$ would be unable to compute the output related to chunk1.

Hence, the introduction of reflux tides ensures that chunks required by each CU return to them at specific steps, enabling complete output computation. Taking CU$_3$ as an example, which corresponds to line 11 in Alg. \ref{alg:Q-stream}, at Step\,3, all CUs transfer chunks while locally replicating them. Consequently, even though other dies do not transfer chunks to CU$_3$ in Step\,3, CU$_3$ retains both chunk1 and chunk5. Following line\,14 in Alg. \ref{alg:Q-stream}, at Step\,4, CU$_3$ participates in reflux tides communication. According to lines 15-16 in Alg. \ref{alg:Q-stream}, during Step\,4, CU$_3$ transfers chunk1 to CU$_2$ and chunk5 to CU$_4$, thereby reintroducing chunk1 and chunk5 to CU$_2$ and CU$_4$, respectively. This operation compensates for the missed computations involving chunk1 and chunk5 on CU$_2$ and CU$_4$ during Step\,2. Therefore, reflux tides not only ensure computational completeness but also minimize communication costs and storage overhead. In this way, MRCA ensures the implementation of DTAttention’s dataflow within a mesh physical topology.



\begin{algorithm}[t]
\setstretch{0.88}
\caption{MRCA Communication Algorithm}\label{alg:Q-stream}
\KwIn{$N$ CUs connected with 1-D mesh topology}
\For{$t = 1$ \textbf{to} $N$}{
    \ParFor{$n \in$ CUs}{
        $src \gets n$\;
        \uIf{$t \leq src < N$}{
            $dest \gets src + 1$, $i \gets src - t + 1$\;
            \texttt{send}($src$, $dest$, chunk[$i$])\;
        }
        \uIf{$1 < src \leq N - t + 1$}{
            $dest \gets src - 1$, $i \gets src + t - 1$\;
            \texttt{send}($src$, $dest$, chunk[$i$])\;
        }
        
        \uIf{$t > \lfloor N/2 \rfloor$}{
            \uIf{$t = \lfloor N/2 \rfloor + 1$}{
                CUs replicate original chunks locally
            }
            \Else{
                \uIf{$t - \lfloor N/2 \rfloor \leq src < t$}{
                    $dest \gets src + 1$,\! $i \gets src + N - t + 1$\;
                    \texttt{send}($src$, $dest$, chunk[$i$])\;
                }
                \uIf{\!$N - t + 1 < src \leq N - t + 1 + \lfloor N/2 \rfloor$}{
                    $dest \gets src - 1$,\! $i \gets src - N + t - 1$\;
                    \texttt{send}($src$, $dest$, chunk[$i$])\;
                }
            }
        }
    }
}
\KwOut{Data orchestration in logical ring on mesh}
\end{algorithm}

\section{Evaluation}\label{sec:Evaluation}
\subsection{Experimental Setup}\label{subsec:Experimental_Setup}
We evaluate the performance of STAR on several representative Transformer models and tasks using the NVIDIA A100 GPU. For computer vision (CV) tasks, we use the latest PVT model \cite{wang2021pyramid} for ImageNet-1k classification, by fine-tuning from the checkpoint pre-trained on ImageNet-21k. For NLP tasks, we select the BERT-base and BERT-large models \cite{devlin2018bert}, which are evaluated across eight tasks from the GLUE benchmark and SQuAD v1.1. Furthermore, we assess GPT-2 \cite{radford2019language}, Bloom-1.7B \cite{le2022bloom}, and Llama7B/13B \cite{touvron2023llama} on language modeling tasks including Wikitext-2, WikiLingua, Wiki-raw, and Winogrande. For each task, fine-tuning is performed on the NVIDIA A100 GPU following token pruning to recover accuracy. {We evaluate models using task-specific metrics: F1 score for SQuAD and STS-B, accuracy for MRPC, RTE, SST-2, QNLI and Winogrande, ROUGE-1 for WikiLingua, and perplexity for Wikitext-2 and Wiki-raw, Top-1 accuracy for ImageNet. For evaluation, we adopt the dense INT16 model as the baseline. We progressively increase the pruning ratio in steps corresponding to a $0.2\%$ accuracy loss (Eq.\eqref{eq:accuracy_loss}) and record the resulting sparse configurations, including the layer-wise top-$k$ settings and the segment sizes. In this way, we obtain a spectrum of sparsity configurations under different controlled levels of accuracy loss. {We evaluate two configurations: standard (0\% loss), aggressive (2\% loss), representing the minimal and maximal performance optimizations.}}   
\begin{equation}
Loss = \Delta \mathcal{M} = \frac{\vert\mathcal{M}_{\text{baseline}} - \mathcal{M}_{\text{STAR}}\vert}{\mathcal{M}_{\text{baseline}}} \times 100\%.
\label{eq:accuracy_loss}
\end{equation}
\vspace{-2mm}



\textbf{Overall Accuracy}. All models are sourced from Pytorch and HuggingFace. The INT16 baselines are derived via post-training quantization. As shown in Table. \ref{tab:accuracy}, the INT16 baseline incurs less than a $1\%$ average accuracy drop from FP16. For reasoning tasks such as MRPC, QNLI and Winogrande, the accuracy degradation caused by INT16 quantization is negligible, typically below $0.5\%$. This aligns with prior work \cite{jacob2018quantization}, which suggests that classification and reasoning tasks, due to their discrete output space and robustness to quantization noise, exhibit a high tolerance for low precision.

For hardware evaluation, we conduct the RTL design for the STAR accelerator and utilized Synopsys DC with TSMC $28$nm CMOS technology, to estimate the area and power consumption of the logic parts. On-chip SRAM bandwidth and power are evaluated with CACTI. For off-chip DRAM, we model memory behaviors with Ramulator \cite{kim2015ramulator} and apply the methodology from \cite{wang2024sofa} to estimate I/O power consumption. We then simulate the RTL with Verilator to extract the actual cycles for each stage, and use this data to build a cycle-level simulator for evaluating end-to-end performance.



For GPU comparisons, we run benchmarks on the A100 platform with the SOTA TensorRT-LLM framework. Execution time is measured by inserting \emph{torch.cuda.synchronize} at run boundaries. For power measurement, we first record the system's idle power with \emph{nividia-smi}, then repeatedly execute workloads to obtain the total power. The dynamic power is derived by subtracting idle power from the total power. 


For evaluation of the spatial architecture, we use ASTRA-sim \cite{won2023astra}, an open-source simulator for distributed machine learning systems, integrated with STAR's cycle-accurate timing model. The simulator is reconfigured with the corresponding NoC communication and DRAM access model, enabling throughput simulation under various workloads.


\begin{table*}[t]
\scriptsize
\renewcommand{\arraystretch}{0.92}
\caption{Accuracy of Diverse Transformer Models with FP16, INT16 and STAR Configurations (S: Standard, A: Aggressive).}\vspace{-1mm}
\begin{tabular}{l|m{0.4cm}<{\centering}m{0.3cm}<{\centering}m{0.3cm}<{\centering}m{0.35cm}<{\centering}m{0.35cm}<{\centering}m{0.5cm}<{\centering}
m{0.42cm}<{\centering}m{0.3cm}<{\centering}m{0.35cm}<{\centering}m{0.35cm}<{\centering}m{0.35cm}<{\centering}m{0.4cm}<{\centering}
m{0.4cm}<{\centering}
m{0.5cm}<{\centering}
m{0.4cm}<{\centering}m{0.58cm}<{\centering}
m{0.4cm}<{\centering}m{0.45cm}<{\centering}
m{0.45cm}<{\centering}m{0.5cm}<{\centering}}
\specialrule{0.12em}{0.5pt}{1pt}
\!\!\textbf{Model} & \multicolumn{6}{c}{BERT-Base} & \multicolumn{6}{c}{BERT-Large} & \multicolumn{1}{c}{GPT-2} & \multicolumn{1}{c}{PVT} & \multicolumn{2}{c}{Bloom1B3} & \multicolumn{2}{c}{LLaMA-7B} & \multicolumn{2}{c}{LLaMA-13B} \\
\hline 
\!\!\textbf{Task}$^{\ddagger}$ &  \!\!\!MRPC  & \!\!RTE & \!\!SST2 & \!\!STSB &  \!\!\!SQuAD\!\!\!\! & QNLI & \!\!MRPC   &  RTE & SST2 & STSB & \!\!SQuAD & QNLI   & \!\!\!\!Wiki2. & \!\!\!ImageN. & \!\!WikiLi. & \!\!WikiRaw & \!\!Wiki2. & \!\!Winog. & \!Wiki2. & \!\!Winog. \\

\!\!\textbf{FP16} & \!\!85.3\%  &  \!\!\!67.1\% & \!\!91.9\% & 83.9 & \!87.4\% & 80.8\%  & \!90.0\% & \!70.1\% & 94.8\% & ~86.5 & 93.2\% & 90.7\% &  \!\!18.3 & 83.6\% & 44.3 & 24.3 & 10.2 & 70.1\% & 8.53 & 75.7\% \\

\!\!\textbf{INT16} & \!\!85.3\%  & \!\!\!67.1\% & \!\!91.9\% & 83.7 & \!87.3\% & 80.8\%  & \!90.0\% & \!70.0\% & 94.8\%  & ~86.5 & 93.2\% &  90.7\% &  \!\!18.7 & 83.4\% &  44.1 & 24.5 & 10.5 & 70.1\% & 8.7 & 75.7\% \\

\!\!\textbf{STAR (S)}\!\! & \!\!85.3\%  & \!\!\!67.0\%  & \!\!91.9\% & 83.7 & \!87.3\% & 80.6\%  & \!89.9\% & \!70.0\% & 94.8\% & ~86.4 & 93.1\% & 90.7\% & \!\!18.7 & 83.4\% & 44.1 & 24.5 & 10.5 & 70.1\% & 8.7 & 75.7\% \\

\!\!\textbf{STAR (A)}\!\! & \!\!83.9\%  & \!\!\!66.7\% & \!\!91.8\% & 83.2 & \!87.1\% & 80.1\%  & \!88.5\% & \!69.2\% & 93.2\% & ~85.1 & 91.4\% & 90.6\% & \!\!18.9 & 81.9\% &43.8 & 25.3 & 10.8 & 69.4\% & 8.9 & 75.4\% \\
\specialrule{0.12em}{0.5pt}{0.5pt}
\end{tabular}
\begin{tablenotes}
\scriptsize
\item \hspace{-4mm} $^{\ddagger}$ MRPC, RTE, SST-2, QNLI, ImageNet, Winogrande are evaluated by accuracy. SQuAD and STS-B are evaluated by F1-score and Pear Correlation, respectively. WikiLingua is evaluated by ROUGE-1. WikiText2 and WikiText Raw are evaluated by perplexity, where lower is better. Standard: 0\% drop vs. INT16; Aggressive: $\leq$2\% drop vs. INT16.
\end{tablenotes}
\label{tab:accuracy}\vspace{-2mm}
\end{table*}

\subsection{Algorithm Performance}\label{subsec:Alg_Evalu}
\textbf{Experimental Settings}. For the objective function of DSE (see Appendix A), the coefficient $\alpha$ controls the weight of the top-$k$ sorting cost, while $\beta$ governs the weight of the SU-FA's exponential cost. We first conduct numerical experiments to define appropriate search ranges for each hyperparameter. During retraining, we apply grid search with successive halving to efficiently identify optimal parameters for each model. Depending on the input-token length and parameter size for different models, $\alpha/\beta$ is set as $0.24$$/$$0.31$ (BERT-B/L), $0.2$$/$$0.24$ (ViT), $0.4$$/$$0.42$ (GPT-2), $0.53$$/$$0.56$ (Bloom-1.7B), and $0.58$$/$$0.63$ (Llama-7B/13B), respectively. We then conduct a search for 200 iterations using each learning rate (1e-1, 5e-2, 1e-3) to identify the optimal tiling configuration.

\textbf{Top-$k$ Hit}. Fig. \ref{fig:layer_error} (a) profiles the hit rate between the predicted top-$k$ by SLZS+SADS, DLZS+SADS and the true top-$k$ during GPT-2 inference. As shown, SLZS+SADS exhibits relatively low hit rates, remaining below 93\% for the top-20\% scenario, largely attributable to estimation errors introduced by leading-zero conversion of both operands. In contrast, DLZS+SADS consistently achieves higher hit rates. For the top-20\% setting, the hit rate exceeds 97\%, with degradation below 90\% observed only in shallow layers under the stricter top-5\% setting. As the network depth increases, the hit rate rises above 95\%, owing to deeper layers extracting higher-level features that yield more distinguishable attention scores. Notably, the higher hit rate in deeper layers helps preserve inference accuracy, as numerical precision near output layers has a greater impact on final performance. 

Fig. \ref{fig:layer_error} (b) further shows that DLZS+SADS incurs at most a 4.8\% accuracy drop without fine-tuning compared to the dense baseline, which is reduced to within 2\% with appropriate fine-tuning. In summary, since DLZS+SADS achieves a high hit rate in deeper layers, which have the most direct impact on the final output, it can effectively preserve the final model performance on downstream tasks. Moreover, with additional fine-tuning, the accuracy loss can be further minimized, making it adaptable to scenarios with different precision requirements.

{\emph{\textbf{TakeAway1}. In top-$k$ settings, decreasing $k$ increases pruning, which helps to reduce computational complexity but may compromise accuracy. STAR provides designers with the flexibility to adaptively balance accuracy and acceleration by adjusting the pruning ratio according to practical needs.}}

\begin{figure*}[t]
\centering
\includegraphics[width=\linewidth]{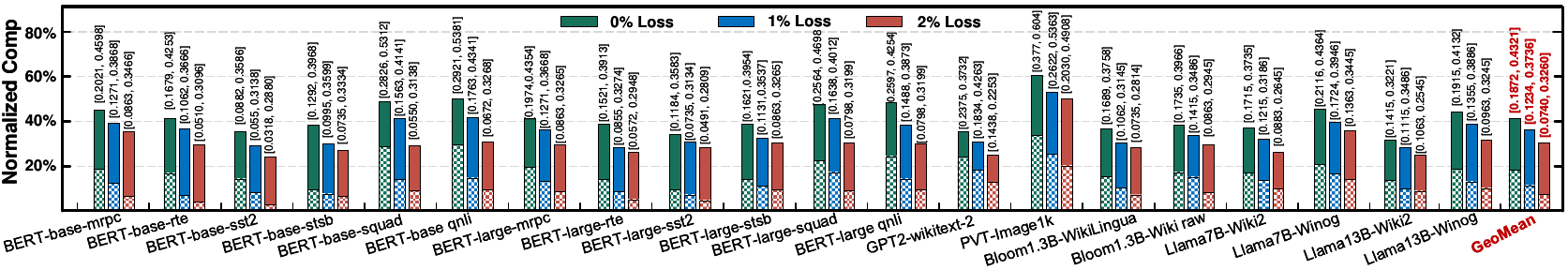}
\caption{Comp reduction by LP with varied loss tolerance. [X,Y] is normalized Comp in Atten and Atten+QKV, respectively.}
\label{fig:overall_complexity_reduction}\vspace{-2mm}
\end{figure*}


\begin{figure}[t]
\centering
\includegraphics[width=0.999\linewidth]{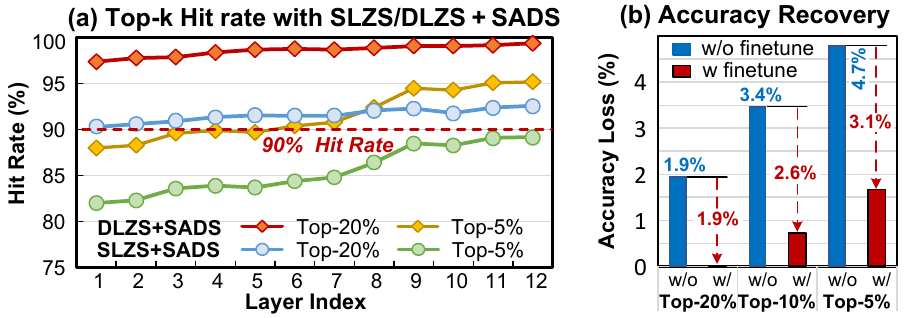}
\caption{{{(a) Layer-wise top-$k$ hit rates of SLZS/DLZS} and (b) End-to-end accuracy analysis of DLZS on GPT-2 WikiText-2.}}
\label{fig:layer_error}\vspace{-2mm}
\end{figure}

\textbf{Ablation Study}. We first set up an ablation study to evaluate the low-complexity benefits of DLZS, SADS, and SU-FA by comparing them with the baseline scheme. The baseline is assumed to employ $4$-bit multiplication during the \emph{pre-compute stage}, vanilla sorting in \emph{top-$k$ stage}, and traditional FA during the \emph{formal-compute stage}. For fairness, we kept the loss of different models within $2\%$. As depicted in Fig. \ref{fig:complexity_reduction}, compared to the baseline, DLZS achieves an average $18\%$ complexity reduction, which is mainly benefits from its multiplier-free design and the use of only half the number of converters. In addition, SADS and SU-FA together contribute an additional $10\%$ reduction, through distributed segmented sorting and the elimination of redundant non-linear operations using top-$k$ hints. Overall, compared to traditional schemes, STAR's software optimizations achieves a $28\%$ reduction in computational complexity for the same token sparsity, thus enabling it to efficiently handle LTPP scenarios. 


{\textbf{Accuracy \& Reduced Complexity Trade-off.} Fig. \ref{fig:complexity_reduction} (b) shows the impact of top-$k$ ratio $\gamma$ on accuracy and reduced complexity (RC) using ViT on ImageNet (CV) and Llama7B on Winogrande (reasoning). Overall, a smaller $\gamma$ results in more aggressive pruning, which may decrease accuracy but increase reduced complexity.  There are some key observations: For CV tasks (ImageNet), accuracy drops noticeably when $\gamma$$<$0.2. In contrast, for text reasoning tasks (Winogrande), the model is more tolerant to pruning, with accuracy degrading evidently only when $\gamma$$<$0.15. This may be because text tasks rely on vital tokens for inference, resulting in higher redundancy. On the other hand, the gains in reduced complexity start to diminish when $\gamma$$<$0.15. This is because overly aggressive pruning harms crucial tokens and limits further complexity reduction. Therefore, to strike a balanced trade-off between accuracy and complexity, we empirically set the top-\textit{k} ratio preferentially within the range of 0.15–0.2. }

\textbf{Overall Complexity Reduction.} Fig. \ref{fig:overall_complexity_reduction} depicts the computation reduction achieved by STAR's sparsity prediction mechanism (LP) across various tasks, using dense models as the baseline and allowing $0\%$, $1\%$, and $2\%$ accuracy loss. As observed, Text-based sentiment tasks (e.g., STT-2 and STSB) exhibit high sparsity, enabling over $90\%$ computation reduction with $1\%$ loss. This is because typically, only a few key terms suffice to capture the sentiment polarity. In contrast, image tasks generally contain more critical information, with lower data redundancy compared to text classification datasets, leading to reduced sparsity. As a result, the computation reduction is limited to around $73\%$ under a $1\%$ loss. Overall, LP achieves significant computation reduction. On average, with accuracy losses of $0\%$, $1\%$, and $2\%$, the LP decreases the attention + QKV computation by $56.8\%$, $62.6\%$, and $67.4\%$, respectively. When focusing on the attention part, computation reduction increases to $81.3\%$, $87.7\%$, and $92.6\%$, respectively. 

\begin{figure}[t]
\centering
\includegraphics[width=\linewidth]{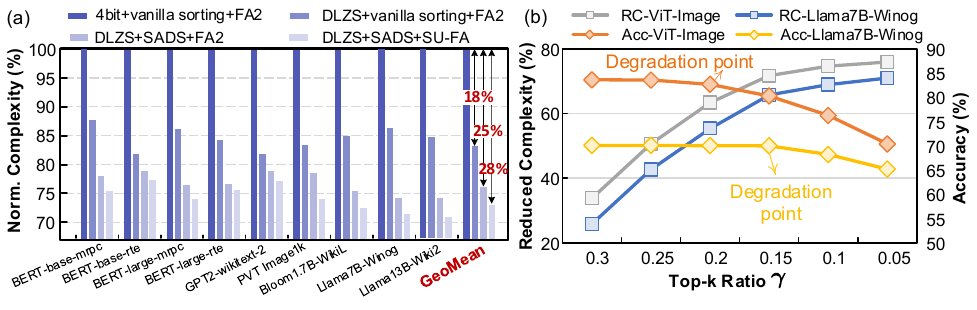}
\caption{(a) Complexity reduction for DLZS, SADS, SU-FA. {(b) Trade-off between accuracy and reduced complexity (RC).}}
\label{fig:complexity_reduction}\vspace{-2mm}
\end{figure}

{\textit{\textbf{TakeAway2}: Textual tasks typically exhibit higher sparsity than vision tasks. Consequently, STAR’s sparsity prediction achieves extremely high computation reduction (over 90\%) in high-redundancy scenarios such as text classification.}}

\subsection{Architecture Evaluation}
\textbf{Throughput over GPU}. Fig. \ref{fig:throughput_gain} shows the throughput gain of STAR over the Nvidia A100 GPU across different tasks. As observed, naively applying the LP mechanism on the GPU results in a limited throughput gain of about $1.08\times$-$1.78\times$. This is because the GPU's coarse-grained parallelism cannot handle token-level fine-grained sparsity, nor can it manage cross-stage on-demand generation and the fine-grained token reallocation of SU-FA. By contrast, the STAR accelerator, with its customized datapath, efficiently pipelines the cross-phase DLZS, distributed SADS, and synergistically updating SU-FA modules, thus achieving a sparse utilization almost triple higher than GPUs. On average, STAR achieves speedups of $6.3\times$$/$$7.0\times$$/$$9.2\times$ with $0\%$$/$$1\%$$/$$2\%$ accuracy loss, respectively. Fig. \ref{fig:Efficiency_gain} (a) illustrates STAR’s effectiveness in reducing memory accesses. Compared to the baseline using vanilla dynamic sparsity, STAR with RASS achieves an average memory access reduction of 23\%. With the incorporation of SU-FA and tiled dataflow, the reduction increases to 79\%.

\begin{figure}[t]
\centering
\includegraphics[width=\linewidth]{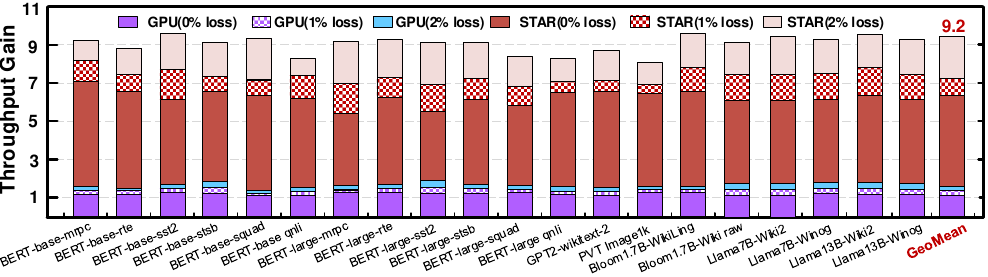}
\caption{{Throughput gain of STAR over LP on A100 GPU.}}
\label{fig:throughput_gain}\vspace{-2mm}
\end{figure}

\begin{figure}[t]
\centering
\includegraphics[width=0.999\linewidth]{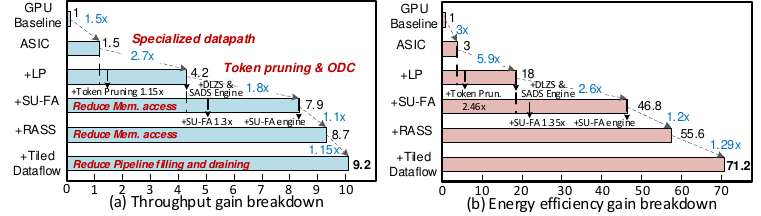}
\caption{Throughput and energy efficiency gain breakdown.}
\label{fig:accleration_breakdown}\vspace{-2mm}
\end{figure}

\textbf{Throughput gain breakdown}. Fig. \ref{fig:accleration_breakdown} shows the breakdown of throughput gain, with the baseline being the execution of the dense model on the A100 GPU. With the dedicated ASIC datapath, STAR is $1.5\times$ faster than the GPU baseline. LP is then applied to find trivial tokens, reducing computation by $3.1\times$. However, the performance only improves by $1.15\times$. This is because dynamic pruning requires pre-computation and top-$k$ to identify redundant tokens, which becomes a bottleneck without dedicated hardware. With the tailored DLZS and SADS engines, performance further improves by $2.7\times$. Similarly, directly applying SU-FA yields only a $1.3\times$ gain due to Max value errors often causing circuit stalls. In contrast, the tailored SU-FA engine can achieve a $1.8\times$ gain. Also, RASS and tiled dataflow together contribute around a $1.27\times$ speedup.

\textbf{Area, Power and Energy:} Fig. \ref{fig:Area_power_breakdown} presents the area and power breakdown of the STAR accelerator implemented in TSMC $28$nm technology. It has a total area of 5.69 mm$^2$ and a power consumption of $949.85$ mW. The LP part (DLZS + SADS) occupies only $18.1\%$ of the area and consumes $14.1\%$ of the power, yet it delivers a $2.8\times$ throughput gain. Fig. \ref{fig:Efficiency_gain} (b) further depicts the energy efficiency improvement over the A100 GPU. Under $0\%$$/$$1\%$$/$$2\%$ accuracy degradation, STAR realizes $49.8\times$$/$$51.6\times$$/$$71.2\times$ efficiency gain, respectively. Fig.\ref{fig:accleration_breakdown} (b) also shows the efficiency gain breakdown. The DLZS and SADS engines provide efficiency gains of $2.58\times$ and $2.3\times$, respectively, while the SU-FA and RASS units together contribute a $3.12\times$ improvement. 


\begin{figure}[t]
\centering
\includegraphics[width=0.92\linewidth]{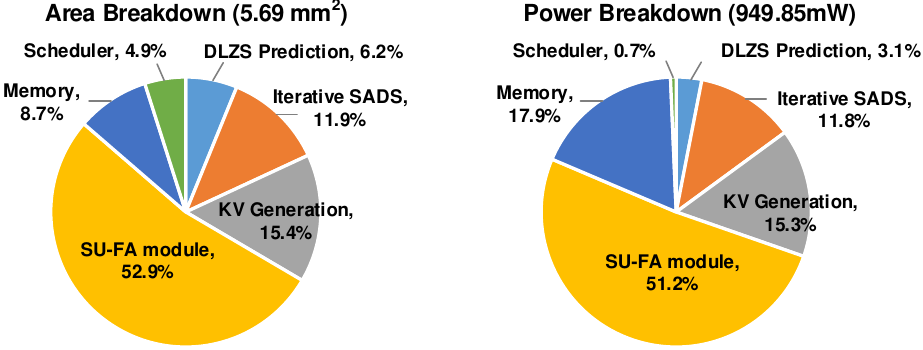}
\caption{Area and power breakdown of STAR accelerator.}
\label{fig:Area_power_breakdown}\vspace{-2mm}
\end{figure}

\begin{figure}[t]
\centering
\includegraphics[width=\linewidth]{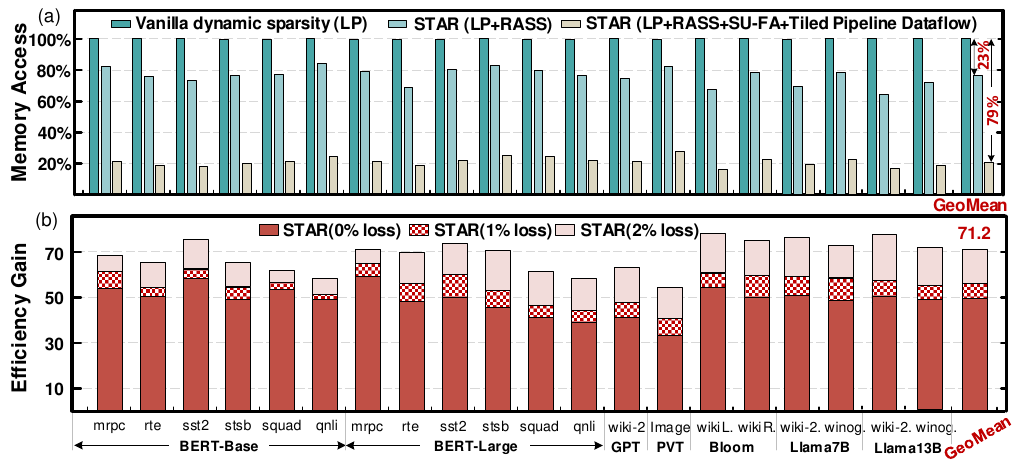}
\caption{{(a) Memory access reduction of STAR. (b) Energy efficiency gain of STAR accelerator over Nvidia A100 GPU.}}
\label{fig:Efficiency_gain}\vspace{-2mm}
\end{figure}

\begin{table}[b]
\footnotesize
\renewcommand{\arraystretch}{1.05}
\caption{{{Summary and comparison with SOTA works.}}}\vspace{-4mm}
\begin{center}
\begin{tabular}{m{2.7cm}<{\raggedright\arraybackslash}|m{1.0cm}<{\centering}m{1.0cm}<{\centering}m{1.0cm}<{\centering}m{1.0cm}<{\centering}}
\specialrule{0.12em}{0.5pt}{1pt}
 & {\textbf{FACT}} & \textbf{Energon} & \textbf{ELSA} & \textbf{STAR} \\
\hline
\!\!\!\!\textbf{Acceleration for}$^\dag$ & {C only} & C only & C only & \textbf{C $\&$ M}  \\
\!\!\!\!\textbf{Optimization level}$^\ast$ & {S. Stg} & S. Stg & S. Stg & \textbf{C. Stg}  \\
\!\!\!\!\textbf{Tech.[nm] \& Freq [Hz]}\!\!\! & {$28$\,/\,500M} & $45$\,/\,1G  & $40$\,/\,1G & $\mathbf{28}$\,/\,1G  \\
\!\!\!\!\textbf{Area[mm$^2$]} & {$6.03$} & $4.20$  & 1.26 & $\mathbf{5.69}$ \\
\!\!\!\!\textbf{Power[W]} & {$0.22$} & $2.72$  & 1.5 & $\mathbf{3.45}$ \\
\!\!\!\!\textbf{Throughput[GOPS]} & {$928$} & $1153$  & 1090 & $\mathbf{24423}$ \\
\!\!\!\!\textbf{Energy Effi.[GOPS/W]}$^{\ddag}$\!\! & {$2754$} & $450$  & $1004$ & $\mathbf{7183}$\\
\!\!\!\!\textbf{Area Effi.[GOPS/mm$^2$]}$^{\ddag}$\!\! & {$154$} & $709$ & $1765$ & $\mathbf{4292}$\\
\specialrule{0.12em}{0.5pt}{0.5pt}
\end{tabular}
\begin{tablenotes}
\footnotesize
{\item \hspace{-4mm} $^\dag$C: Computation. M: Memory access; $^\ast$ S. Stg: Single stage; C. Stg: Cross stage;~$^{\ddag}$Scaled to the 28nm technology and 1.0V with $f$ $\propto$ $s$, power (core) $\propto$ $(1/s)(1.0/Vdd)^2$ as \cite{wang2024sofa,wang2023efficient}, where $s=$Tech\,/\,28nm.}
\end{tablenotes}
\end{center}
\label{tab:hardware_compara}\vspace{-4mm}
\end{table}

{Fig. \ref{fig:Cache_Analysis}(a) shows the throughput performance under a single-core architecture when adopting the LP+SU-FA+RASS+Tiled dataflow design compared with the baseline across different SRAM capacities. The baseline corresponds to the conventional attention computation paradigm without softmax tiling. The results indicate that with 316 kB of SRAM, the LP+SU-FA+RASS+Tiled dataflow design achieves saturated throughput, while the baseline remains memory-bound due to frequent memory accesses and thus fails to reach peak throughput.}

{Fig. \ref{fig:Cache_Analysis}(b) illustrates the performance gains achieved under a 5$\times$5 spatial architecture when applying the DRAttention+MRCA optimization scheme. The baseline is configured without these optimizations as well as SU-FA and RASS. In the spatial architecture, memory constraints become more pronounced due to the increased competition for memory bandwidth among multiple compute units. With 412 kB of SRAM, the baseline achieves only 3 TOPS, whereas DRAttention+MRCA reaches 24.1 TOPS—12$\times$ higher—thanks to its efficient dataflow and enhanced data reuse opportunities.}

\begin{figure}[t]
\centering
\includegraphics[width=\linewidth]{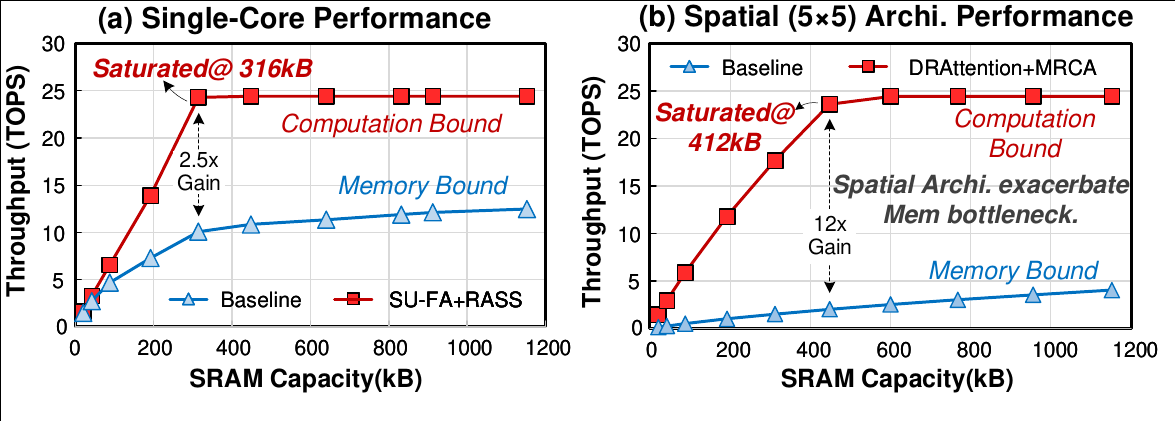}\vspace{-2mm}
\caption{Impact of SRAM size on average single-core throughput with/without memory-access optimizations under (a) single-core with 256GB/s DRAM bandwidth (b) {a multi-core configuration with 512GB/s total (shared) DRAM bandwidth, i.e., 20.5 GB/s effective bandwidth per core.}}
\label{fig:Cache_Analysis}\vspace{-2mm}
\end{figure}

{\textit{\textbf{TakeAway3}: Without customized dataflow and communication algorithms (e.g., DRAttention) to mitigate memory-access overhead, employing a multi-core architecture can actually degrade throughput due to contention for memory resources.}}

\subsection{Comparison with Existing Accelerators}
Sanger \cite{lu2021sanger}, FACT \cite{qin2023fact}, Energon \cite{zhou2022energon}, SpAtten \cite{wang2021spatten} and ELSA \cite{ham2021elsa} are SOTA Attention DS accelerators. However, they all prioritize computation optimization and overlook memory access, which becomes the dominant bottleneck as computation efficiency improves. While SpAtten reduces memory traffic via head pruning and Energon considers compute–memory balance, both suffer from task and model dependency, lacking of generality. Overall, prior designs fail to jointly optimize computation and memory access under sparsity.

In contrast, STAR employs a holistic FlashAttention-like scheme that tiles all DS stages and exploits sorting information for cross-stage optimization. {Table \ref{tab:hardware_compara} compares the characteristics and metrics of FACT, Energon, ELSA and STAR. STAR's energy efficiency (with cross-stage compute and memory optimizations) reaches 7183 GOPS/W, which is $2.6\times$, $15.9\times$ and $7.2\times$ (tech normalized \cite{wang2024sofa,wang2023efficient}) higher than FACT, Energon and ELSA (single-stage computation-only optimization). This improvement results from the fine-grained data flow enabled by collaborative cross-stage optimization, effectively reducing off-chip memory accesses. Additionally, STAR achieves $4292$ GOPS/mm$^2$ in area efficiency, which is $27.1\times$, $6.1\times$ and $2.4\times$ greater than FACT, Energon and ELSA, respectively. The area efficiency gain is primarily due to algorithm-hardware co-optimization for low complexity. }

\subsection{Performance of Spatial Architecture}

We first conduct an ablation study to evaluate the effectiveness of the \textit{DTAttention} dataflow and the customized \textit{MRCA} communication algorithm in improving STAR's performance on spatial architectures. Evaluation configurations are summarized in Table \ref{tab:wafer_config}. {The baseline adopts a Ring-Attention (ICLR'23) \cite{liu2023ring}, where KV tensors are transmitted among compute units following a ring topology, without any topology-aware or communication-efficiency optimizations.} 

{As shown in Fig. \ref{fig:wafer_performance} (a), under the 5$\times$5 scenario,\textit{ DRAttention} achieves an average $3.1\times$ throughput gain, primarily due to its strategy of transmitting smaller Query tensors. This allows communication latency to be better overlapped with computation, thereby improving compute unit utilization. However, this logic-level optimization is inherently constrained by the mismatch between logical and physical topologies. With the integration of MRCA, throughput further increases by $3.6\times$, attributed to efficient communication orchestration that mitigates tail latency from boundary compute units. As the system scales to a $6\times6$ mesh (Fig. \ref{fig:wafer_performance} (b)), the gain from \textit{DRAttention} diminishes to $3\times$, whereas \textsc{MRCA} achieves a higher gain of $4.2\times$. This is because the larger-scale system results in more limited average memory bandwidth, emphasizing \textsc{MRCA}'s superior capability in reducing contention and tail latency.}

\begin{table}[b]
\footnotesize
\renewcommand{\arraystretch}{1.05}
\caption{Spatial archi. configuration parameters.}\vspace{-4mm}
\begin{center}
\begin{tabular}{l|cc}
\specialrule{0.12em}{0.5pt}{1pt}
\!\!\textbf{Modules} & \textbf{Parameters} & \!\!\textbf{Configurations}\!\! \\
\hline
\!\!\multirow{4}{*}{Logic Part}\!\! 
&\!\!Logic Die Area\!\! & 324mm$^2$\\
& \!\!Die Router \!\! &  Input-queued Architecture \\
& \!\!Routing Algorithm \!\! &  Dimension-Order \\
& \!\!Die-to-Die Interconnect \!\! & 250GB/s, 20ns, 1.0pJ/bit \\
\hline
\!\!\multirow{1}{*}{DRAM Part}\!\! 
&\!\! HBM-die Area / Data Rate \!\!& 240\,mm$^2$ / @2Gbps/pin \\
(HBM\,2) &\!\!Access Bandwidth\!\! & 512GB/s, 100ns, 6.0pJ/bit  \\
\specialrule{0.12em}{0.1pt}{0.1pt}
\end{tabular}
\end{center}
\label{tab:wafer_config}\vspace{-2mm}
\end{table}


Fig. \ref{fig:wafer_performance} (c) compares the performance of diverse spatial architectures integrated with diverse computing units. The baseline is assumed to adopt the same NVDLA compute architecture as Simba \cite{shao2019simba}, utilizing SIMD-based multiple parallel vector MAC units to perform dense attention computations, referred to as \emph{Spatial-Simba}. In contrast, the spatial architectures using Spatten \cite{wang2021spatten} and STAR as the compute unit are referred to as \emph{Spatial-Spatten} and \emph{Spatial-STAR}, respectively. As depicted in Fig. \ref{fig:wafer_performance} (c), under 5$\times$5 scenario, \emph{Spatial-Spatten} archives an average $6.7\times$ gain while \emph{Spatial-STAR} improves this gain further to $20.1\times$. For \emph{Spatial-Spatten}, its benefits mainly stem from the computation acceleration for token sparsity. Unfortunately, the heavy IO burden still hinders its performance gain, especially for the spatial architecture scenarios, where memory access heavily relies on the transfer by NoC routers. Heavy traffic increases the contention in the NoC network, which in turn degrades the throughput. In contrast, strategic cross-stage tiling in \emph{Spatial-STAR} efficiently alleviates the IO burden, making it well-suitable for spatial architecture. {When scaling up to a $6\times6$ architecture, as depicted in Fig. \ref{fig:wafer_performance} (d), the performance gap caused by the I/O optimization capability becomes further amplified. The average throughput gain of \textit{Spatial-Spatten} decreases to $5.6\times$, whereas that of \textit{Spatial-STAR} further increases to $22.8\times$.}

{\textbf{Scalability}. STAR handles long sequences by combining spatial unfolding (DRAttention) and temporal unfolding (SU-FA, Tiling), enabling scalability to arbitrarily long inputs. For ultra-long sequences, it only requires assigning additional workload to each STAR unit, which can process them by extending the number of time steps.}


{\textbf{Load-Balance Handling.} The DRAttention dataflow inherently mitigates load imbalance through uniform workload partitioning and overlapped compute–communication scheduling. 1) Each STAR unit receives an equal-sized query sub-block and performs local attention while simultaneously transmitting and receiving sub-blocks along the mesh. This pipelined execution hides minor timing variations among units, ensuring near-uniform utilization. 2) Further, the MRCA employs a reflux phase that synchronizes boundary tiles and compensates for straggling nodes, effectively reducing tail latency. These mechanisms together maintain balanced progress even under small variations in compute or link latency. }


\begin{figure}[t]
\centering
\includegraphics[width=\linewidth]{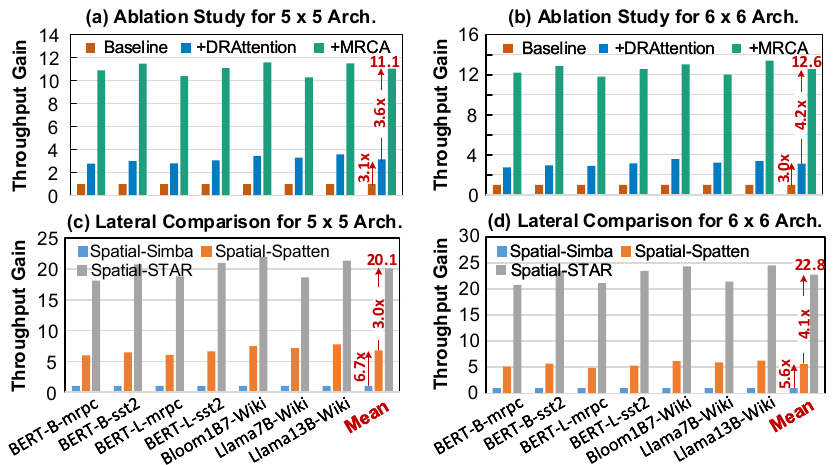}
\caption{{(a)(b) Ablation studies for DRAttention and MRCA. (c)(d) Lateral comparison for diverse computing units.}}
\label{fig:wafer_performance}\vspace{-2mm}
\end{figure}

\section{Related Works and Discussion}\label{sec:Related_Works}
\emph{\textbf{Efficient Transformer Accelerators}}. Numerous works~\cite{ham20203, ham2021elsa, lu2021sanger, wang2021spatten, qu2022dota, zhou2022energon, qin2023fact,you2023vitcod, li2020ftrans, shen2022salo, wang2025pade, zhao2024hardware,wang2025beta} aim to improve Transformer inference efficiency. However, most of these efforts focus on alleviating attention's quadratic computation, by static sparsity~\cite{you2023vitcod, li2020ftrans, shen2022salo}, dynamic sparsity~\cite{ham20203, ham2021elsa, lu2021sanger, wang2021spatten, qu2022dota, zhou2022energon, qin2023fact}, and hybrid sparsity~\cite{zhao2024hardware}. Unfortunately, in many cases, memory access is the de facto bottleneck in both power and end-to-end latency, particularly in LTPP scenarios. To address this, STAR jointly optimizes computation and memory access, offering substantial improvements over prior works. STAR adopts a cross-stage holistic optimization strategy. This unique methodology enables cross-stage tiling, facilitating fine-grained dataflow tiling that accelerates inference while significantly reducing off-chip memory access.


\emph{\textbf{Neural Network Accelerators with Sparsity}}. A variety of ASIC and FPGA accelerators leverage sparsity to optimize neural network inference performance~\cite{hojabr2021spaghetti,asgari2020alrescha,wang2018low,wang2025mcbp}. However, most existing work focuses on pre-trained static sparse weights, while STAR utilizes LP to predict dynamic sparsity on-the-fly, particularly exploiting the argmax approximation of softmax, which requires active detection. This makes traditional static zero-based sparsity methods ineffective. While some works have targeted activation sparsity~\cite{jang2021sparsity} and both weight and activation sparsity~\cite{wu2023highlight}, they are all based on near-zero sparsity, failing to address the top-$k$ sparsity that STAR targets.



\section{Conclusion}\label{sec:Conclusion}
This paper presents STAR, a scalable Transformer accelerator for LTPP, enabled by a cost-efficient log-domain predictor, distributed sorting and simplified FlashAttention. To further scale across spatial architectures, we propose a customized dataflow and a tailored communication algorithm. Evaluated on 5$\times$5 and 6$\times$6 spatial architectures,  STAR delivers $20.1\times$ and $22.8\times$ throughput gains over the baseline.

\bibliographystyle{IEEEtran}
\footnotesize
\bibliography{IEEEabrv,main}

\vfill

\end{document}